\documentclass[12pt, reqno]{amsart}
\usepackage{amsmath,amssymb,pdflscape,bm,xr}
\usepackage{amsfonts,natbib}
\usepackage{tikz}
\usepackage{tikz-network}
\usepackage{graphicx,epsfig,setspace}
\usepackage[margin=1in]{geometry}

\numberwithin{equation}{section}
\DeclareMathOperator*{\argmin}{arg\,min}

\begin{document}

\newcommand{\cov}{\textnormal{Cov}}
\newcommand{\var}{\textnormal{Var}}
\newcommand{\diag}{\textnormal{diag}}
\newcommand{\plim}{\textnormal{plim}_n}
\newcommand{\dum}{1\hspace{-2.5pt}\textnormal{l}}
\newcommand{\ind}{\bot\hspace{-6pt}\bot}
\newcommand{\co}{\textnormal{co}}
\newcommand{\tr}{\textnormal{tr }}
\newcommand{\fsgn}{\textnormal{\footnotesize sgn}}
\newcommand{\sgn}{\textnormal{sgn}}
\newcommand{\fatb}{\mathbf{b}}
\newcommand{\fatp}{\mathbf{p}}
\newcommand{\trace}{\textnormal{trace}}
\newcommand{\Eta}{\textnormal{H}}

\newtheorem{dfn}{Definition}[section]
\newtheorem{rem}{Remark}[section]
\newtheorem{cor}{Corollary}[section]
\newtheorem{thm}{Theorem}[section]
\newtheorem{lem}{Lemma}[section]
\newtheorem{notn}{Notation}[section]
\newtheorem{con}{Condition}[section]
\newtheorem{prp}{Proposition}[section]
\newtheorem{pty}{Property}[section]
\newtheorem{ass}{Assumption}[section]
\newtheorem{ex}{Example}[section]
\newtheorem*{cst1}{Constraint S}
\newtheorem*{cst2}{Constraint U}
\newtheorem{qn}{Question}[section]

\onehalfspacing

\title[Structural Sieves]{Structural Sieves}
\author[Konrad Menzel]{Konrad Menzel\\New York University}
\date{\emph{Date:} October 2021.}

\begin{abstract}This paper explores the use of deep neural networks for semiparametric estimation of economic models of maximizing behavior in production or discrete choice. We argue that certain deep networks are particularly well suited as a nonparametric sieve to approximate regression functions that result from nonlinear latent variable models of continuous or discrete optimization. Multi-stage models of this type will typically generate rich interaction effects between regressors (``inputs") in the regression function so that there may be no plausible separability restrictions on the ``reduced-form" mapping form inputs to outputs to alleviate the curse of dimensionality. Rather, economic shape, sparsity, or separability restrictions either at a global level or intermediate stages are usually stated in terms of the latent variable model. We show that restrictions of this kind are imposed in a more straightforward manner if a sufficiently flexible version of the latent variable model is in fact used to approximate the unknown regression function.\\[4pt]


\noindent\textbf{JEL Classification:} C14,C25,C43,C45,C50\\
\textbf{Keywords:} Artificial Neural Networks, Deep Learning, Nested Models, Production Functions, Multinomial Choice
\end{abstract}

\maketitle

\section{Introduction}


Artificial Neural Networks (ANN) have been extremely successful at solving certain statistical tasks involving the interpretation of sensory data and replication of human cognition. This paper pursues the question whether a potential advantage relative to other flexible approximation devices may carry over to certain types of economic data.  We consider plausible generative models of production and discrete choice that have a latent structure satisfying shape or other qualitative constraints which do not directly translate into manageable restrictions on a ``reduced form" in terms of the manifest (observable) variables of the model.

We propose a flexible, ``deep" modeling approach for estimation of economic models. We envision a scenario in which reality is complex but modular in a way that is best captured by a, possibly nonlinear, latent variable model. That is, mappings transforming observable inputs into outputs can be disaggregated into simpler components with a simpler structure. We focus in particular on the role of shape, sparsity and separability restrictions  that are imposed globally, or at intermediate stages of that transformation. These properties are in general not inherited by a composition of multiple such stages (``layers") consisting of multiple parallel units (``neurons"), but can be imposed as sign or exclusion restrictions in estimation approaches that replicate this modular structure.

In contrast to general-purpose sieves to estimate reduced-form relationships between the observed variables nonparametrically, our approach consists in using a parametric model for estimation which can then be made arbitrarily flexible. We see several benefits to such an approach - for one, if economic behavior that is best described in terms of a latent variable structure, a reduced form will typically display interaction effects in forcing variables if the relationship is not fully linear. A model with a similar latent variable structure may more readily reproduce such interaction effects even at a fairly low degree of approximation, especially if these are disciplined by a fairly low-dimensional latent factor structure. Furthermore, common qualitative model restrictions - such as monotonicity, convexity, separability, or sparsity - are usually imposed on economic primitives but have no analog in the resulting reduced form. We show that shape restrictions of this kind can be easily imposed as sign or exclusion restrictions in the deep generating model and also greatly restrict its expressivity, resulting in superior theoretical performance. Finally, we can also report ancillary aspects or predictions of the estimated model to help interpret and provide context to the main empirical results.

A particular feature of our approach is that we impose qualitative shape restrictions on the function of interest, concavity or quasi-concavity, and monotonicity, which are motivated by economic theory. These restrictions substantially reduce the complexity of the class of functions that can be represented by the network and therefore improve our ability to approximate the function flexibly. Furthermore, convexity may also reduce computational challenges from multiple local minima of the loss function when training the neural network.

The dimension-dependent optimal rates of nonparametric estimation derived in \cite{Sto80} impose an absolute constraint on how well an otherwise unconstrained statistical relationship between multiple variables can be estimated. With that in mind, our results speak to three key scenarios, depending on sample size and the researcher's confidence in the modular structure of the underlying DGP and the importance of shape restrictions.
\begin{enumerate}
\item The researcher may have high confidence in qualitative (shape, sign, sparsity, separability) restrictions on economic primitives but not necessarily the reduced form, which can then be directly imposed in the approach proposed in this paper.
\item In ``data-poor" settings when key aspects of the data generating process are only poorly identified from the data, our approach will favor low-dimensional aspects according to a modular structure rather than based on a generic linear sieve. This allows the researcher to choose the direction of likely inductive biases in favor of an interpretable, plausible model.
\item Finally, we also give universal approximation rates that do not depend on a specific modular structure and give asymptotic guarantees in ``data rich" settings.
\end{enumerate}
Our approach therefore seeks to combine a parametric ``structural" modeling philosophy with a nonparametric approach, where we do not assume a ``correct" specification of the data generating process but an approximation that can be made arbitrarily flexible depending on the available amount of data. Through a nonparametric lens, it is known that the choice of estimators for nonparametric model components does not affect first-order asymptotic properties of regular estimators (see e.g. \cite{New94}), however we are interested in settings when the data may not be sufficiently rich for these formal conclusions to be taken at face value. Rather, we propose a ``regularization path" along a sequence of parsimonious parametric models that are adapted to the economic structure of the problem but also have the usual properties of a nonparametric sieve. Results can also be interpreted ``parametrically" by reporting ``projections" of components of the full model onto a nested, lower-dimensional version of the model in addition to a main effect of functional of primary interest.

We will develop our main ideas for nested CES models which are particularly suited for problems with convexity constraints. It would be possible to consider different sieves for modular problems without convexity, however at a greater challenge for controlling statistical complexity. A theory for estimation of nested separable nonparametric models using B-splines has been derived by \cite{HMa07}.

\subsection{Related Literature}
Our setup has obvious parallels with popular methods in deep learning, most importantly multilayer feedforward neural networks (MFNN) and deep Boltzmann machines (DBM) consisting of multiple hidden layers, each of which processes inputs from the preceding layers into a vector of outputs. One key difference of our approach is that we consider settings in which these nested transformations are not mechanical but involve decisions by economic agents who may also anticipate their effect on subsequent layers of the process. We also allow for unobservables to enter at each stage of this process rather than treating the nested model as deterministic.

Various authors, most importantly \cite{HMa07}, \cite{ESh16}, \cite{MPo16}, \cite{KKr17}, \cite{BKo19}, and \cite{SHi20} have shown that deep network architectures have a superior performance in uncovering compositional models of nested layers of smooth functions that satisfy certain separability or sparsity conditions. Compositionality of this type is common in latent variable models which have historically played an important role in describing economic decisions, see e.g. \cite{McF74}, \cite{Tra09}, or \cite{Hec79}. Also, key concepts in economic models - economic activity, human or physical capital, etc. - are often not directly observable as a scalar quantity but are typically inferred as indices or proxies constructed from measurements or components of that latent concept. Deep learning has so far been most fruitfully applied in image and video processing which can also be viewed as latent variable problems, where surfaces defining an object in space can typically not all be seen at the same time on an image or may be occluded by other objects.

This paper explores the applicability of techniques from deep learning to economic problems, when observable outcomes of economic activity are best described as a result of multi-stage decision or production processes and when it is impractical to model the intermediate stages of that process explicitly e.g. due to the lack of direct measurements or the need to specify additional model components parametrically or nonparametrically. For an overview over recent developments in deep learning, see \cite{GBC16} and \cite{ZLLS20}. Recent work by \cite{Yar17} and \cite{FLM19} derives inferential results for semiparametric inference using deep neural networks with a ReLU (rectified linear units) activation function.


The paper is organized as follows: Section 2 gives the general framework for the model and estimation, section 3 analyzes nested models of production as approximations for a general technology, and section 4 gives rates for approximating a discrete choice model with general dependence among taste shocks using multi-layer cross-nested Logit models. Section 5 discusses network architectures to impose additional qualitative constraints on the approximating network. Section 6 gives our main asymptotic result regarding the rate of nonparametric estimation with and without additional shape restrictions, section 7 concludes.

\section{General Framework}

We consider the problem of flexible estimation of a reduced-form relationship between variables $\mathbf{X}:=(X_{1},\dots,X_{K})'$ (``explanatory variables", ``covariates", or ``inputs") and $\mathbf{Y}:=(Y_{1},\dots,Y_{M})'$ (``outcomes", ``outputs"). This could be for the purpose of prediction, causal inference, or in the context of a structural model. We also let $\mathbf{Z}:=(\mathbf{Y}',\mathbf{X}')'$.

We assume that the researcher observes a sample of $n$ i.i.d. units $i$ from a distribution with joint p.d.f.
\[f_0(\mathbf{y},\mathbf{x}) = f_{Y|X}(\mathbf{y}|\mathbf{x})f_X(\mathbf{x})\]
assuming that the conditional p.d.f. is also well-defined for $\mathbf{x}$. We allow for the possibility of missing data, where certain components of $\mathbf{Y}$ and/or $\mathbf{X}$ may not be observed for some units in the sample. We assume that the object of interest is a function depending on that distribution,
\[\mu_0(\mathbf{z}):=\mu(\mathbf{z};f_0)\]
Here we will primarily consider cases in which $\mu_0(\mathbf{z})$ denotes a conditional or unconditional expectation or probability for an event in $\mathbf{Y}$ given $\mathbf{X}$.

We assume that $\mu_0$ is defined as the minimizer of expected loss
\begin{equation}\label{tau_0_def}\mu_0:=\arg\min_{\mu}\mathbb{E}[\ell(\mu,\mathbf{Z})]\end{equation}
where the random vector $\mathbf{Z}$ is distributed according to $f_0(\mathbf{z})$. Following \cite{FLM19} we assume that the loss function $\ell(\mu,\mathbf{z})$ is Lipschitz
\[|\ell(\mu,\mathbf{z})-\ell(\mu',\mathbf{z})|\leq C_l|\mu(\mathbf{z})-\mu'(\mathbf{z})|\]
for some constant $C_l<\infty$, and satisfies
\[c_1\mathbb{E}[(\mu(\mathbf{Z})-\mu_0(\mathbf{Z}))^2]\leq\mathbb{E}[\ell(\mu,\mathbf{Z})-\mathbb{E}[\ell(\mu_0,\mathbf{Z})]\leq c_2\mathbb{E}[(\mu(\mathbf{Z})-\mu_0(\mathbf{Z}))^2]\]
for finite constants $c_2>c_1>0$.

For the purposes of this paper, the first leading case is that of a conditional expectation function, $\mu_0(\mathbf{x}):=\mathbb{E}[Y|\mathbf{X}=\mathbf{x}]$ for scalar $Y$, which satisfies (\ref{tau_0_def}) with $\ell(\mu,\mathbf{z}):=(y-\mu(\mathbf{x}))^2$. A second case of interest is that of a conditional choice probability where $\mathbf{Y}$ takes values in a finite set $\{y_1,\dots,y_M\}$ and $\mu(y,\mathbf{x}):=\mathbb{P}(Y=y|\mathbf{X}=\mathbf{x})$ is the conditional probability of $Y=y$ given $\mathbf{X}$. One loss function satisfying (\ref{tau_0_def}) for this problem is $\ell(\mu,\mathbf{z}):=-\sum_{m=1}^M\dum\{Y=y_m\}\log\mu(y_m,\mathbf{x})$.

Nonparametric estimation of conditional mean or distribution functions is known to be subject to a curse in dimensionality in the number of covariates and/or outcomes, see \cite{Sto80}, which is inherent in the estimation problem and not the particular technique that is used for estimation. The key challenge here is that in the absence of additional separability restrictions, a flexible model for this relationship would have to account for any possible interaction effects between functions of two or more components of $\mathbf{X}$.

The approach put forward in this paper does not sidestep or remedy this challenge, however we propose a nonlinear sieve that (a) incorporates common shape constraints on the underlying economic model (most importantly convexity), and (b) specifically targets the interaction effects that would result from standard models of economic decisions and optimization. We argue that when observable data results from nested nonlinear models, then separability of the reduced form will be the exception and not the rule, and ``deep" architectures may have an advantage at replicating these interaction effects with a smaller number of parameters.

To appreciate this point, consider a linear index model with a scalar outcome variable $Y_i$ and regression function
\[\mu(\mathbf{x}):=\mathbb{E}[Y_i|\mathbf{X}_i=\mathbf{x}] \equiv G(\mathbf{x}'\boldsymbol\beta)\]
with coefficient $\beta\in\mathbb{R}^K$ and link function $G:\mathbb{R}\rightarrow\mathbb{R}$ that is twice continuously differentiable. The cross-partial derivatives of this model,
\[\frac{\partial^2}{\partial x_k\partial x_l}\mu(\mathbf{x})=\beta_k\beta_lG''(\mathbf{x}'\boldsymbol\beta)\]
are generally non-zero, unless the function $G(\cdot)$ is affine. As this simple example illustrates, a single nonlinear transformation (``activation") may be sufficient to mask any separability properties that may have been satisfied in preceding stages of this model. However those interaction effects are also tightly constrained by the simple parametric structure of this model, so an estimation approach exploiting that index structure may in fact estimate the conditional mean function at a much faster rate.


A similar point was established formally for a class of nonlinear compositional or hierarchical interaction models by \cite{BKo19} and \cite{SHi20}. The potential benefits of deeper, rather than shallow, network architectures to approximate compositional functions have recently been analyzed by \cite{ESh16} and \cite{MPo16}. Specifically, if the model in $K$ regressors is a composition of a bounded number of functions of at most $d^*\leq K$ variables at each step, estimation by a sufficiently deep network can achieve estimation at a nonparametric rate depending on $d^*$ rather than $K$. This suggests that nested models may have an advantage at leveraging sparsity or separability restrictions at intermediate stages to achieve faster convergence rates and mitigate the curse of dimensionality.

As a second key feature, analyzing the data-generating process as a latent variable model naturally incorporates missing or mismeasured data into estimation. This obviously includes the cases where the available data are complete or are thought to conform to a ``missing at random" assumption (see \cite{Rub76}), but also situations where missing data is at the heart of the problem, including imperfect measurement (\cite{JGo75}, \cite{CHS10}), self-selected samples (\cite{Hec79},\cite{APo93}), and causal inference (\cite{Ney23},\cite{Rub78}). More generally, such an approach can be adapted to situations in which data quality is uneven or observed variables are only proxies the relevant economic concept.

\subsection{Generative Model: Nested Decisions}

We develop a technique for solving and estimating models involving a large number of nested - discrete or continuous - intermediate decisions to approximate complex statistical relationships between inputs and outputs. One key distinguishing feature of our approach is that ``activations" at intermediate stages incorporate not only states in past layers that are carried forward mechanically, but also constitute the result of choices by an optimizing agent who anticipates outputs in subsequent layers, iterating future states backward.


This results in an multilayer perceptron (MLP) with possibly nonlinear activation functions in which intermediate states can be fed both forward and backwards, so that the resulting graph is not necessarily acyclic. For the purposes of estimation, these intermediate stages are latent and not directly observable to the researcher. We then show that such a model can be made arbitrarily flexible by increasing the number of units (``nodes") in each hidden stage (``layer"). 

For expositional clarity, we focus on two prototypes for such generative models with latent discrete or continuous decisions, which may be adapted or combined flexibly. The first model concerns a model for a production technology, where initial inputs are transformed into intermediate goods, which in turn serve as inputs in subsequent intermediate and final stages of production. Specifically, we assume that production takes place in $S+1$ stages (``layers"), where at the $s$th layer there are separate technologies (``neurons") to produce intermediate goods $k=1,\dots,K_s$ for $s=0,\dots,S$.

The technology for producing intermediate good $k$ is given by the production function
\[w_k^{(s)} = \tilde{F}_k^{(s)}(w^{(s-1)}_1,\dots,w^{(s-1)}_{K_{s-1}})\]
where $w^{(s-1)}_l$ denotes the quantity of the $l$th intermediate good from the $(s-1)$ stage employed in the production of the $k$th intermediate good at stage $s$. We show below that the optimal (cost-minimal) production plan can be characterized recursively as the cost-minimization problem at neuron $k$ in layer $s$, with input prices $\pi_1^{(s-1)},\dots,\pi_{K_{s-1}}^{(s-1)}$ given by marginal cost of production at stage $s-1$, and given desired output level $v_k^{(s)}$ determined by factor demand at stage $s$. The values of solution of $v_k^{(s)},\pi_k^{(s)}$ resulting from this constrained optimization problem are determined recursively via the activation mappings
\begin{eqnarray}
\label{forward_iteration}\pi_k^{(s)}&=&\tilde{\phi}_k^{(s)}\left(\pi_1^{(s-1)},\dots,\pi_{K_{s-1}}^{(s-1)};v_1^{(s)},\dots,v_{K_s}^{(s)}\right)\\
\label{backward_iteration}v_k^{(s)}&=&\tilde{\psi}_k^{(s)}\left(\pi_1^{(s)},\dots,\pi_{K_s}^{(s)};v_1^{(s+1)},\dots,v_{K_{s+1}}^{(s+1)}\right)
\end{eqnarray}
that take states $\left(h_l^{(s-1)}\right)_{l=1}^{K_{s-1}}:=\left(\pi_l^{(s-1)},v_l^{(s-1)}\right)_{l=1}^{K_{s-1}}$ and $\left(h_l^{(s+1)}\right)_{l=1}^{K_{s+1}}:=\left(\pi_l^{(s+1)},v_l^{(s+1)}\right)_{l=1}^{K_{s+1}}$, respectively, as inputs and produce outputs $\left(h_k^{(s)}\right)_{k=1}^{K_s}$.

The second model is a model of nested discrete decisions where at the $k$th neuron in the $s$th layer an agent chooses among $K_{s+1}$ discrete nests in the subsequent layer. There are no flow payoffs, but each terminal node in the top layer is associated with a random utility where the joint distribution of taste shocks corresponds to that for a cross-nested Logit (CNL) with that nesting structure (see \cite{Vov97} and \cite{WKo01}).


This model can be reparametrized as sequential choice among nests starting at the bottom layer, given the inclusive (continuation) values $v_k^{(s)}$ for the $k$th nest in the $s$th layer. We also denote the unconditional probability of reaching the $k$th nest in layer $s$ with $\pi_k^{(s)}$. Inclusive values are determined by iterating the activation mappings
\begin{eqnarray}
\nonumber v_k^{(s)}&=&\tilde{\psi}_k^{(s)}\left(\pi_1^{(s)},\dots,\pi_{K_s}^{(s)};v_1^{(s+1)},\dots,v_{K_{s+1}}^{(s+1)}\right)
\end{eqnarray}
backwards from the $S$th (top) layer to the root node of the graph, and the conditional choice probabilities for the nests can be obtained by iterating forward over
\begin{eqnarray}
\nonumber\pi_k^{(s)}&=&\tilde{\phi}_k^{(s)}\left(\pi_1^{(s-1)},\dots,\pi_{K_{s-1}}^{(s-1)};v_1^{(s)},\dots,v_{K_s}^{(s)}\right)
\end{eqnarray}
from the root node. For the CNL specification the mappings $\psi_k^{(s)}(\cdot),\phi_k^{(s)}(\cdot)$ are available in closed form.



\subsection{Approximating Model: Artificial Neural Network}

The primitive functions $F_k^{(s)}(\cdot)$ in the generative model are generally not known and are typically not nonparametrically identified (see e.g. \cite{HMa07} for feedforward networks). We therefore  do not aim to estimate intermediate activation functions but are primarily interested in estimating the reduced form $f_0(\mathbf{y},\mathbf{x})$ given shape restrictions on the primitive functions. To this end we propose an artificial neural network with the capacity to approximate the activation functions $\psi_k^{(s)},\phi_k^{(s)}$ flexibly while incorporating qualitative constraints on $F_k^{(s)}(\cdot)$.

We construct a network of $\tilde{S}+1$ layers, where layer $s$ has $\tilde{K}_s$ neurons for $s=0,\dots,\tilde{S}$. For neuron $k$ in the $s$th layer we choose the production function (or flow utility for a network of nested discrete decision) from a parametric family,
\[\tilde{F}_k^{(s)}(w_1,\dots,w_{\tilde{K}_{s-1}}) = \tilde{f}\left(w_1,\dots,w_{\tilde{K}_{s-1}};\boldsymbol\theta_k^{(s)}\right)\]
where the vector $\boldsymbol\theta_{k}^{(s)}$ consists of parameters  $\boldsymbol\beta_k^{(s)}=(\beta_{k1}^{(s)},\dots,\beta_{kK_{s-1}}^{(s)})$ governing the effect of the preceding layer on activation of neuron $k$ in layer $s$, and potentially additional shape parameters $\boldsymbol\varrho_k^{(s)}$.

\begin{figure}
\SetCoordinates[yLength=1,xLength=1,zLength=1]
\begin{tikzpicture}
\Vertices[Math]{opt_based_vertices.csv}
\Edges{opt_based_edges.csv}
\end{tikzpicture}
\caption{Nested production functions (left) and parametrization of the optimal production plan (right).}
\label{fig:opt_based_network}
\end{figure}
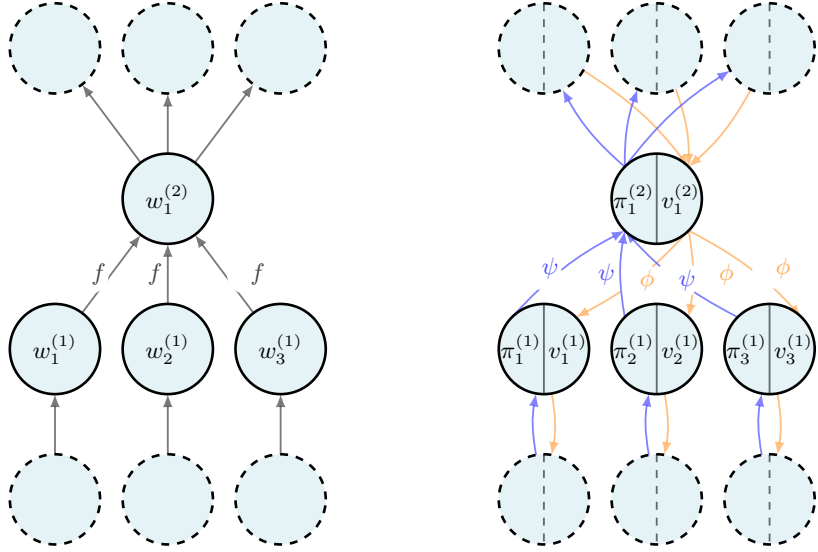

We then solve the constrained optimization problem analogous to the generative model given state variables $v_1^{(s+1)},\dots, v_{\tilde{K}_{s+1}}^{(s+1)}$ and $\pi_1^{(s-1)},\dots,\pi_{\tilde{K}_{s-1}}^{(s-1)}$ to obtain the activation functions
\begin{eqnarray}
\nonumber \pi_k^{(s)}&=&\phi_k^{(s)}\left(\pi_1^{(s-1)},\dots,\pi_{K_{s-1}}^{(s-1)};v_1^{(s)},\dots,v_{K_s}^{(s)}\right)\\
\nonumber&\equiv&\phi_k^{(s)}\left(\pi_1^{(s-1)},\dots,\pi_{K_{s-1}}^{(s-1)};v_1^{(s)},\dots,v_{K_s}^{(s)};\boldsymbol\theta_{k-}^{(s)} \right)\\
\nonumber v_k^{(s)}&=&\psi_k^{(s)}\left(\pi_1^{(s)},\dots,\pi_{K_s}^{(s)};v_1^{(s+1)},\dots,v_{K_{s+1}}^{(s+1)}\right)\\
\nonumber&\equiv&\psi_k^{(s)}\left(\pi_1^{(s)},\dots,\pi_{K_s}^{(s)};v_1^{(s+1)},\dots,v_{K_{s+1}}^{(s+1)};\boldsymbol\theta_{k+}^{(s)} \right)
\end{eqnarray}
for some collection of parameters $\boldsymbol\theta_{k-}^{(s)},\boldsymbol\theta_{k+}^{(s)}$ from the adjacent layers, $s-1$ and $s+1$.


Such a structure with $S$ layers and activation functions $\psi_k^{(s)},\phi_k^{(s)}$ defines a class $\mathcal{H}_{\mathbf{K}}^S$ of neural networks that is indexed by the free parameters $\boldsymbol\theta$. We denote the activations of the hidden units that are consistent with the recursions (\ref{forward_iteration}) and (\ref{backward_iteration}) with $h_k^{(s)}(\boldsymbol\theta)$.



\subsection{Estimation and Prediction}

Given the sample $\mathbf{z}_1,\dots,\mathbf{z}_n$, we can train this network, where we identify $\mu(\mathbf{z}_i)\equiv\mu(\mathbf{z}_i;\boldsymbol\theta)$ with activations $h_{ki}^{(s)}(\boldsymbol\theta)$ of certain neurons, in most cases in the top and/or bottom layer of the network. Specifically, we estimate $\mu_0(\mathbf{z})$ by minimizing the average training error,
\begin{equation}\label{tau_hat_def1}\hat{\mu}\in\arg\min_{\mu\in\mathcal{H}_{\mathbf{K}}^{S}\\ \|\mu\|_{\infty}\leq 2B}\frac1n\sum_{i=1}^n\ell(\mu,\mathbf{z}_i)\end{equation}
for some bound $B<\infty$. Here, minimization over the class $\mathcal{H}_{\mathbf{K}}^{S}$ is equivalent to minimization over the parameter $\boldsymbol\theta$. Note that training error in (\ref{tau_hat_def1}) is the empirical analog of expected population loss in
(\ref{tau_0_def}), the criterion defining the target $\mu_0$. The trained network can then be used to compute $\hat{\mu}(\mathbf{z})$ for arbitrary values of $\mathbf{z}$.

Our main theoretical results concern statistical properties of $\hat{\mu}$ as an estimator for $\mu_0$, and recommendations for the choice of the number and size of hidden layers. The minimization problem in (\ref{tau_hat_def1}) can be solved adapting commonly used algorithms for conventional feedforward networks, including stochastic gradient descent, or Adam. We argue in Appendix \ref{sec:backprop_app} that the gradient can be approximated using recursive application of the chain rule even when the graph is not acyclic.

The hidden layers of this model are generally not identified in the sense that the minimum in (\ref{tau_hat_def1}) may be attained (exactly or to an approximation) at multiple values of $\boldsymbol\theta$. We do not address this issue explicitly in this paper, rather the main object of interest for the purposes of this paper is a reduced-form relationship characterizing the joint distribution of the observable variables $\mathbf{z}_i$. A structural interpretation of the hidden layers may require additional restrictions and normalizations, see also \cite{HMa07} for a discussion for the case of nonparametric estimation of nested regression functions.

\subsection{Examples}

\subsubsection{Multiple Measurements} Consider a model for production of outputs $Y_i^*$ from inputs $X_i$ where we do not observe $Y_i^*$ directly, but rather a collection of measurements $Y_i:=(Y_{1i},\dots,Y_{Mi})$. For example there is no direct agreed upon measure for human capital, rather the term captures groups of distinct skills and attitudes, see e.g. \cite{CHS10}. In a typical setting, we observe parental investments, indicators of educational (e.g. test scores) and economic achievement (e.g. labor force status or income) and want to determine the effects of various educational investments or interventions.

\begin{figure}
\SetCoordinates[yLength=1,xLength=1,zLength=1]
\begin{tikzpicture}
\Vertices{mimic_vertices.csv}
\Edges{mimic_edges.csv}
\end{tikzpicture}
\caption{Multiple indicators, multiple causes (MIMIC) model for a latent variable $Y^*$ and measurements $Y_1,\dots,Y_4$.}
\end{figure}
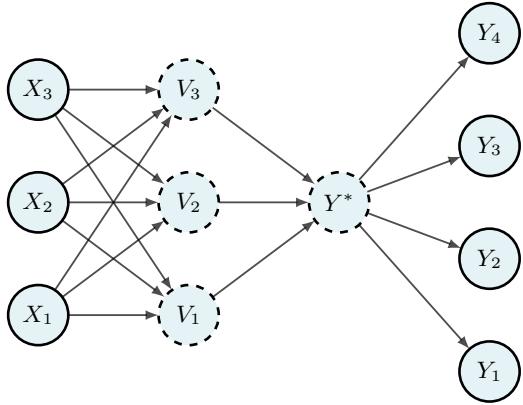

\cite{JGo75} proposed the parametric Multiple Indicators and Multiple Causes (MIMIC) model with a single hidden variable to estimate such a production process. \cite{CHS10} analyzed a semi-parametric model for human capital production with two latent factors, cognitive and non-cognitive skills which could in principle be further disaggregated into more distinct latent skills to explain the link between childhood investments and testing or economic performance.

Our approach involves a disaggregate description of a technology where we the number of latent factors can be chosen flexibly, and we may observe only some inputs directly as quantities, and possibly prices or cost shifters for some others. This approach involves a trade-off between the predictive performance of the model and interpretability, which should be resolved depending on the ultimate objective and the functionals/parameters $\mu$ used to summarize this technology.

\subsubsection{Causal Inference} Suppose that we are interested in the causal effect of a binary intervention (``treatment") $D\in\{0,1\}$ on a scalar outcome variable $Y$. Using the Neyman-Rubin potential outcomes framework (see e.g. \cite{IRu15}), each unit $i$ is associated with two possible outcomes, $Y(1)$ with the intervention $D=1$, and $Y(0)$ in the absence of an intervention, $D=0$. For any unit in the sample, only the potential outcome corresponding to the realized intervention, $Y=Y(D)=(1-D)Y(0) + DY(1)$, is observed. In this setup, the unit-level treatment effect $\Delta:=Y(1)-Y(0)$ can not be determined from the observable data, so that the problem of causal inference can be viewed as a missing-data problem.

Suppose that we have a sample of $n$ units for which we observe all relevant covariates $X$, the realized treatment $D$, and the outcome $Y=Y(D)$ Furthermore, assume that we observe additional variables $\mathbf{Z}$ that are conditionally independent of $Y(0),Y(1)$ given $X$, but not of $D$ (under the ignorability assumption for observational designs, $D$ itself would meet that requirement). We can then represent this problem in our framework with a bivariate outcome vector $(Y(1),Y(0))'$, where the network architecture can be constrained such that $Z$ serves as an indirect input determining $D$, but is excluded from the latent factor model determining potential outcomes $(Y(0),Y(1))$.

\begin{figure}
\SetCoordinates[yLength=1,xLength=1,zLength=1]
\begin{tikzpicture}
\Vertices{treatment_vertices.csv}
\Edges{treatment_edges.csv}
\end{tikzpicture}
\caption{Instrumental Variable Model for the Causal Effect of $D$ on $Y$.}
\end{figure}
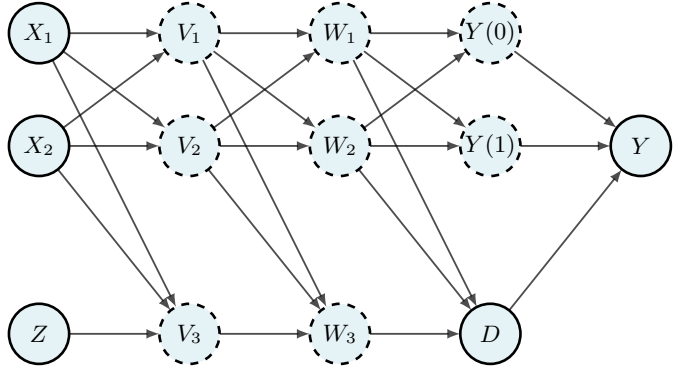

After training the MLP relating $X_i,\mathbf{Z}_i$ to $Y_i,D_i$, we can simulate from the estimated distribution of $Y_i,D_i$ given $X_i$ and different values of $\mathbf{Z}_i=z$. In particular we can evaluate the conditional expectation of $Y_i(1)-Y_i(0)$ among the compliers according to that estimated distribution. We may in principle also compute other functionals of the estimated distribution that are not nonparametrically identified, however these will in general not be consistent estimates of their population counterparts but vary with the parametrization and the choice of starting values during training of the MLP.





\section{Nested Production Functions}

Following the seminal work by \cite{Bec64} and \cite{Bec65}, decisions on fertility, investment in education, and other forms of economic activity have been cast as a problem of optimal allocation of resources towards efficient home production.  Nested models of production have been considered by \cite{McF78b}, \cite{PWa87}, or \cite{PWa75} in the context of household production, where \cite{PRu95} established a general approximation property for nested Constant Elasticity of Substitution (CES) function with regards to the cost functions.

Misspecified production functions can easily lead to inaccurate policy conclusions or obscure economically relevant features of real-world processes. As an important example, \cite{CHe07} and \cite{CHS10} show in their seminal work on skill formation that an extension of a single-skill model with a single period of production to two skill types (cognitive and non-cognitive) and multiple successive periods of investments and accumulation uncovers richer dynamic complementarities in the skill-formation process, and matches important empirical facts about skill formation that can't be properly explained by the simpler model. Our approach would allow to disaggregate latent production stages into a larger number of distinct skills (``intermediate goods") and stages of production. However our focus is on approximating a reduced form mapping between observable inputs and outputs, so we do not analyze identification of intermediate technologies, which was central to their contribution.

\begin{figure}
\SetCoordinates[yLength=1,xLength=1,zLength=1]
\begin{tikzpicture}
\Vertices{skill_formation_vertices.csv}
\Edges{skill_formation_edges.csv}
\node[text width=2cm, anchor=west, right] at (-0.6,-2.3) {$t=1$};
\node[text width=2cm, anchor=west, right] at (1.9,-2.3) {$t=2$};
\node[text width=2cm, anchor=west, right] at (4.4,-2.3) {$t=3$};
\node[text width=2cm, anchor=west, right] at (6.9,-2.3) {$t=4$};
\node[text width=2cm, anchor=west, right] at (9.4,-2.3) {$t=5$};
\end{tikzpicture}
\caption{5-Period Production of skills $v_t^{c},v_t^{nc}$ from investments $x_t^c,x_t^{nc}$ with measurements $y_t^c,y_t^{nc}$ based on
\cite{CHe07}.}
\end{figure}
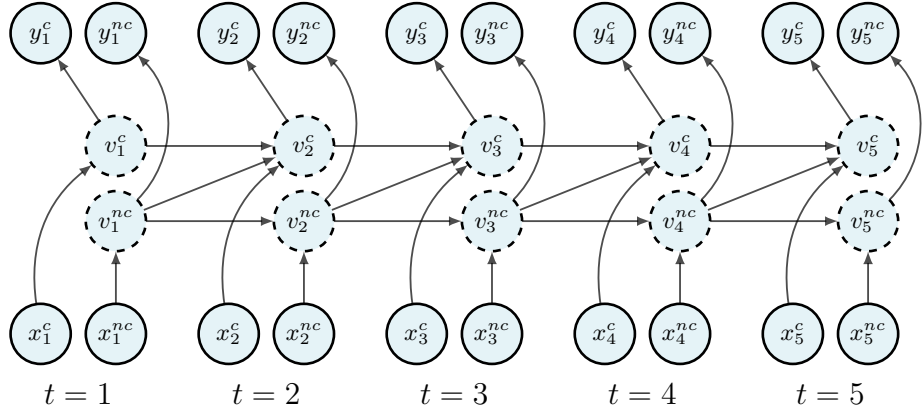

We consider a model of production with a large number of unobserved intermediate goods to approximate a general technology to approximate a mapping from partially observed factor quantities and/or prices to multiple outputs. There are $K$ basic inputs $\mathbf{x}=\left(x_1,\dots,x_K\right)'$ and $M$ final outputs $\mathbf{y}=(y_1,\dots,y_M)'$. We describe the technology as the production possibility set $\mathcal{Y}_0\subset\mathbb{R}^{M+K}$, where a pair of $(\mathbf{y}',\mathbf{x}')'$ is interpreted as a transformation of inputs $-\mathbf{x}$ to produce outputs $\mathbf{y}$. We also let
\[\mathcal{Y}_0(\mathbf{x}):=\left\{\mathbf{y}:(\mathbf{y}',\mathbf{x}')'\in\mathcal{Y}_0\right\}\]
denote the set of feasible outputs given inputs $\mathbf{x}$. We maintain the following assumptions on the technology:

\begin{ass}\label{technology_ass}\textbf{(Technology)} $\mathcal{Y}_0$ is a closed, convex subset of $\mathbb{R}^{M+K}$ and satisfies the following: (a) $0\in\mathcal{Y}_0$. (b) Free disposal: if $(\mathbf{y}',\mathbf{x}')'\in\mathcal{Y}_0$, then $(\mathbf{y}'-\mathbf{t}_1,\mathbf{x}'-\mathbf{t}_2)'\in\mathcal{Y}_0$ for any $\mathbf{t}_1,\mathbf{t}_2\geq0$. (c) Monotonicity: if $(\mathbf{x},\mathbf{y})$ is on the boundary of $\mathcal{Y}_0$, then $(\mathbf{y}'+\mathbf{t}_1,\mathbf{x}'-\mathbf{t}_2)'\notin\mathcal{Y}_0$ for any $\mathbf{t}_1,\mathbf{t}_2\geq0$.
\end{ass}

While these assumptions are common in classical production theory, they are also restrictive. Taken together, parts (1) and (3) preclude the existence of fixed costs of production. Convexity of the production possibility set also restricts returns to scale to be nonincreasing. A commonly imposed, weaker version of (3) only assumes that input sets are convex from below, which is still sufficient for duality theory to apply (see e.g. \cite{McF78b}). For our approximation results, weakening parts (1) and (3) would require recentering of intermediate inputs, thus introducing additional free parameters and complicating the approximating argument. We therefore maintain the stronger set of assumptions throughout.

We can characterize the convex hull of $\mathcal{Y}_0(\mathbf{x})$  in terms of its support function,
\[\mu_0(\mathbf{u},\mathbf{x}):=\sup\left\{\langle\mathbf{y},\mathbf{u}\rangle:(\mathbf{y},\mathbf{x})\in\mathcal{Y}_0\right\},\hspace{0.5cm}
\mathbf{u}\in\mathbb{R}_+^M\]
For the case of a single output, $\mu_0(y,\mathbf{x})$ corresponds to the usual production function, and for the general case $M\geq1$, the support function $\mu_0(\mathbf{e}_m,\mathbf{x})$ evaluated at the $m$th unit vector $\mathbf{e}_m$ yields the highest achievable level for the $m$th output given inputs $\mathbf{x}$. We generally assume efficient production, $\langle\mathbf{y},\mathbf{y}\rangle
=\mu_0(\mathbf{y},\mathbf{x})$, however we do not explicitly model selection among feasible output combinations on the frontier.

\subsection{Nested CES Production Functions}

To approximate the production function associated with $\mathcal{Y}_0$, we consider a production network of the following form: production takes place in $S$ stages, where at stage $s$, the $K_{s-1}$ intermediate inputs $\mathbf{w}^{(s-1)}$ produced during the preceding stage are transformed into $K_{s}$ intermediate outputs $\mathbf{w}^{(s)}=(w_1^{(s)},\dots,w_{K_s}^{(s)})'$.

At each stage, the transformation of intermediate inputs into intermediate outputs is described by Constant Elasticity of Substitution (CES) production functions. Specifically, we define
\begin{eqnarray}\label{CES_activation_fct}
F\left(w_{1},\dots,w_{K};\boldsymbol\theta_k^{(s)}\right)
&=&\left[\sum_{l=0}^{K}\left(\beta_{kl}^{(s)}w_{lk}^{(s-1)}\right)^{\varrho}
\right]^{\frac{\tau}{\varrho}}
\end{eqnarray}
for parameters $\boldsymbol\theta_k^{(s)}:=(\beta_{k1}^{(s)},\dots,\beta_{kK_{s-1}}^{(s)},\varrho_k^{(s)},\tau_k^{(s)})'$,
and assume that output at the $k$th neuron in the $s$th layer is given by
\begin{eqnarray}
\nonumber w_k^{(s)}&=&F_k^{(s)}(w_{1k}^{(s-1)},\dots,w_{K_{s-1}k}^{(s-1)})\\
\nonumber &=&F\left(w_{1k}^{(s-1)},\dots,w_{K_{s-1}k}^{(s-1)};\boldsymbol\theta_k^{(s)}\right)
\end{eqnarray}
Here, $w_{lk}^{(s-1)}$ is the quantity of intermediate good $w_l^{(s-1)}$ committed to the production of the intermediate output $w_k^{(s)}$, and we set $w_0^{(s-1)}\equiv1$ for each $s=1,\dots,S$, which can be thought of as a non-discretionary input.

Intermediate goods may be rival or non-rival as inputs for future stages of production, and we assume that they cannot be acquired from outside sources, but can only be produced using this technology. Note that the stage technologies in (\ref{CES_activation_fct}) include the identity $w_k^{(s)} = w_l^{(s-1)}$ for any $k,l$, so that intermediate outputs could be passed through multiple stages of production unaltered.

We then identify the intermediate inputs at the first stage with the basic inputs, so that any feasible production plan must satisfy the resource constraints
\[w_l\geq\left\{\begin{array}{lcl}\sum_{k=1}^{K_1}w_{lk}^{(0)}&\hspace{0.5cm}&\textnormal{if good }l\textnormal{ is rival}\\ \max_{k=1,\dots,K_{1}}w_{lk}^{(0)}&&\textnormal{if good }l\textnormal{ is non-rival}\end{array}\right.\]
Similarly, the intermediate outputs at the last stage are identified with the final outputs,
\[y_m\leq w_m^{(S)}\hspace{0.3cm}\textnormal{for any }m=1,\dots,M=:K_S\]
Furthermore, any feasible production plan must satisfy the resource constraints for intermediate goods,
\begin{equation}\label{prod_res_const}w_l^{(s-1)}\geq\left\{\begin{array}{lcl}\sum_{k=1}^{K_{s-1}} w_{lk}^{(s-1)}&\hspace{0.5cm}&\textnormal{if good }l\textnormal{ is rival}\\ \max_{k=1,\dots,K_{s-1}}w_{lk}^{(s-1)}&&\textnormal{if good }l\textnormal{ is non-rival}\end{array}\right.\end{equation}
For simplicity, the remainder of this paper will only consider the case in which all intermediate goods are rival.

We denote the resulting technology with $\mathcal{Y}^{S}_{\mathbf{K}}\subset\mathbb{R}_+^{M+K}$, that is the set of input/output vectors $(\mathbf{y}',\mathbf{x}')'$ that can be achieved via a feasible production plan with intermediate production functions (\ref{CES_activation_fct}), satisfying constraints (\ref{prod_res_const}), where the vector $\mathbf{K}=(K_1,\dots,K_S)'$. We denote the corresponding support function with
\[\mu_{\mathbf{K}}^S(\mathbf{u},\mathbf{x}):=\sup\left\{\langle\mathbf{y},\mathbf{u}\rangle:(\mathbf{y},\mathbf{x})\in\mathcal{Y}_{\mathbf{K}}^S\right\},\hspace{0.5cm}
\mathbf{u}\in\mathbb{R}_+^M\]

The feasible set $\mathcal{Y}^{S}_{\mathbf{K}}\subset\mathbb{R}_+^{M+K}$ can be described recursively as follows: at stage $s$, we let $\mathcal{W}^{(s)}$ denote the feasible set of quantities for the intermediate goods from stages $s'=0,\dots,s$, that is the set of values $(\mathbf{w}^{(s)},\mathbf{w}^{(s-1)},\dots,\mathbf{w}^{(0)})$ such that there exist $\tilde{w}_{kl}^{(s-1)}$ for $k=1,\dots,K_s$ and $l=1,\dots,K_{s-1}$ such that $F_k^{(s)}(w_{k1}^{(s-1)},\dots,w_{kK_{s-1}}^{(s-1)})\geq w_k^{(s)}$ and $(\mathbf{w}^{(s-1)}+\tilde{\mathbf{w}}^{(s-1)},\mathbf{w}^{(s-2)},\dots,\mathbf{w}^{(0)})\in\mathcal{W}^{(s-1)}$, where $\tilde{\mathbf{w}}^{(s-1)}=\sum_{k=1}^{K_s}\left(\tilde{w}_{k1}^{(s-1)},\dots,\tilde{w}_{kK_{s-1}}^{(s-1)}\right)'$. Iterating from $\mathcal{W}^{(0)}:=\mathbb{R}^K$ we obtain the set $\mathcal{W}^{(S)}$ of feasible combinations of inputs and intermediate goods for all $S$ stages, so that we can define $\mathcal{Y}^{S}_{K_1,\dots,K_S}$ as the intersection of $\mathcal{W}^{(S)}$ with $\mathbf{R}^K\times\{0\}\times\dots\times\{0\}\times\mathbb{R}^M$, projected on its $K+M$ nontrivial coordinates.

By construction, $\mathcal{W}^{(s)}$ is convex for each $s$. Since $\mathcal{Y}^{S}_{K_1,\dots,K_S}$ is the intersection of that set with a linear subspace, it is also convex. Furthermore, if $\varrho_k^{(s)}>-\infty$ for all $s,k$, the efficient frontier of $\mathcal{W}^{(s)}$ is continuously differentiable to any order, implying that the boundary of the feasible set $\mathcal{Y}^{S}_{K_1,\dots,K_S}$ is also arbitrarily smooth. If $\varrho_k^{(s)}=-\infty$ for some stages and intermediate goods, we will instead consider limits as $\varrho_k^{(s)}\rightarrow-\infty$ for arguments relying on differentiability of the boundary.

In sum our approximating model consists of nested CES technologies, where the elasticity of substitution $\frac{1}{1-\varrho}$ and the degree of homogeneity $\tau$ are allowed to differ across intermediate technologies. Our main claim in this section is that this technology of nested CES aggregators is sufficient to approximate any technology $\mathcal{Y}_0$ satisfying our main assumptions.

\begin{prp}\label{CES_prod_universal_prp}\textbf{(Universal Approximator)} Suppose the production possibility set $\mathcal{Y}_0$ satisfies Assumption \ref{technology_ass}. Then for any $\delta>0$ and compact rectangular set $\mathcal{C}\subset\textnormal{int}\mathbb{R}_{+}^{K+M}$, there exists an approximating technology $\mathcal{Y}_{\mathbf{K}}^S$ constructed via a production network of $S=2$ stages and intermediate technologies of the CES form in (\ref{CES_activation_fct}) such that
\[\sup_{(\mathbf{u},\mathbf{x})\in\mathcal{C}}\left|\mu_{\mathbf{K}}^S(\mathbf{u},\mathbf{x})-\mu_0(\mathbf{u},\mathbf{x})\right|\leq\delta\]
where the number of free parameters is $KK_1$, and
\[K_1 = c(R,K+M)\delta^{-\frac{K+M-1}{2}}\]
for a constant $c(R,K+M)$ that only depends on the dimension $M$ and diameter $R$ of $\mathcal{C}$.
\end{prp}

See the appendix for a proof. The argument is in fact constructive and  establishes that only two stages of production are needed, where basic inputs are perfect complements ($\varrho_k^{(1)}=-\infty$ for each $k$) in the production of the intermediate outputs $\mathbf{w}^{(1)}$, which in turn are perfect substitutes ($\varrho_k^{(2)}=1$ for all $k$) in the production of final outputs $\mathbf{y}$. This nested production function spans a polytope of production possibilities in $\mathbb{R}_+^{K+M}$. The vertices spanning that polytope correspond to production plans involving only a single intermediate output, with all remaining components of $\mathbf{w}^{(1)}$ equal to zero. Since the technology for intermediate output $w_k^{(1)}$ has constant returns to scale up to output $\beta_{k0}^{(1)}$, the second-stage technology then convexifies that vertex set. We can then approximate $\mathcal{Y}_0$ arbitrarily well by adding a sufficient number of appropriately chosen vertices to that polytope. It is also instructive to compare the approximation rate to the approximation bound for RELu approximation of smooth functions in \cite{Yar17}, where the assumption of convexity results in a rate comparable to that for the case of weak derivatives up to order 2.

\subsection{Optimal Production Plan} We next derive the activation functions $\phi(\cdot),\psi(\cdot)$ implied by maximizing behavior in the production model. Given the technology $\mathcal{Y}^{S}_{K_1,\dots,K_S}$ and input prices $\mathbf{p}=(p_1,\dots,p_K)$ we assume that outputs $\mathbf{\hat{y}}$ are produced using a cost-optimal, feasible production plan. That is,
\[(\mathbf{\hat{y}}',\mathbf{\hat{x}}')' :=\arg\min_{\mathbf{x}}\left\{\mathbf{p'x}:(\mathbf{\hat{y}}',\mathbf{x}')'\in\mathcal{Y}^{S}_{K_1,\dots,K_S}\right\}\]
We characterize the optimal production plan at each of the $S$ stages both in terms of quantities of intermediate inputs employed in each intermediate technology, as well as the implied price of each intermediate good.

At stage $s$, the optimal production plan for $w_k^{(s)}\equiv v_k^{(s)}$ units of the $k$th intermediate good given implied input prices $\pi_0^{(s-1)},\dots,\pi_{K_{s-1}}^{(s-1)}$ for the intermediate outputs from the $(s-1)$st stage is the solution to the cost-minimization problem
\[\min_{v_{1k}^{(s-1)},\dots,v_{K_{s-1}k}^{(s-1)}}\sum_{l=0}^{K_{s-1}}\pi_l^{(s-1)}v_{lk}^{(s-1)}\hspace{0.5cm}\textnormal{subject to }
F_k^{(s)}(v_{0k}^{(s-1)},\dots,v_{K_{s-1}k}^{(s-1)}) = v_k^{(s)}\]
where the non-discretionary input is held fixed at $v_{0k}^{(s-1)}=1$.

The value of this constrained optimization problem is given by the cost function
\[C_k^{(s)}(\boldsymbol{\pi}^{(s-1)};v_k^{(s)})\equiv C\left(\boldsymbol{\pi}^{(s-1)};v_k^{(s)};\boldsymbol\theta_k^{(s)}\right)\]
where
\begin{eqnarray}
\nonumber C\left(\boldsymbol{\pi};v;\boldsymbol\theta\right)&:=&\pi_0
+\left[\sum_{l=1}^{K}\left(\frac{\pi_l^{\varrho}}{\beta_{l}}\right)^{\frac1{\varrho-1}}\right]^{\frac{\varrho-1}{\varrho}}
\left(v^{\frac{\varrho}{\tau}}-\beta_{0}^{\varrho}\right)^{\frac1{\varrho}}
\end{eqnarray}
Our framework treats quantities and implicit prices of the intermediate goods as latent (hidden) variables which are determined recursively by solving the cost-minimization problem for stages $s=1,\dots,S$. Specifically, the price of intermediate good $k$ corresponds to the marginal cost of production of an additional unit of the good,
\begin{eqnarray}
\nonumber \pi_k^{(s)}&=& \frac{\partial}{\partial v_k^{(s)}}C_k^{(s)}(\boldsymbol\pi^{(s-1)};v_k^{(s)})\\
\nonumber&=:&\phi_k^{(s)}\left(\boldsymbol{\pi}^{(s-1)};\mathbf{v}^{(s)};\boldsymbol\theta_k^{(s)}\right)
\end{eqnarray}
where we define
\begin{eqnarray}
\label{prod_update_price}
\phi_k^{(s)}\left(\boldsymbol\pi;\boldsymbol{v};\boldsymbol\theta\right)
&:=&\left[\sum_{l=1}^{K}\left(\frac{\pi_l^{\varrho}}{\beta_{l}}\right)^{\frac1{\varrho-1}}\right]^{\frac{\varrho-1}{\varrho}}
\left(v_k^{\frac{\varrho}{\tau}}-\beta_{0}^{\varrho}\right)^{\frac1{\varrho}-1}
\frac{v_k^{\frac{\varrho}{\tau}-1}}{\tau}
\end{eqnarray}
Given prices, optimal factor demand for intermediate input $k$ for intermediate technology $l$ at stage $s+1$ is
\[v_{kl}^{(s)}=\left(\frac{\pi_k^{(s)}}{\beta_{lk}^{(s+1)}}\right)^{\frac1{\varrho_l^{(s+1)}-1}}
\left[\sum_{j=1}^{K_{s}}\left(\frac{(\pi_j^{(s)})^{\varrho_l^{(s+1)}}}{\beta_{lj}^{(s+1)}}\right)^{\frac1{\varrho_l^{(s+1)}-1}}
\right]^{-\frac1{\varrho_l^{(s+1)}}}
\left((v_l^{(s+1)})^{\frac{\varrho_l^{(s+1)}}{\tau_l^{(s+1)}}}-(\beta_{l0}^{(s+1)})^{\varrho_l^{(s+1)}}\right)^{\frac1{\varrho_l^{(s+1)}}}
\]
so that total factor demand for intermediate good $k$ is
\begin{eqnarray}
\nonumber v_{k}^{(s)}&:=&\sum_{l=1}^{K_{s+1}}v_{kl}^{(s)}=:\psi_k^{(s)}\left(\boldsymbol{\pi}^{(s)};\mathbf{v}^{(s+1)};\boldsymbol\theta^{(s+1)}\right)
\end{eqnarray}
where we define
\begin{eqnarray}
\label{prod_update_input}
\psi_k^{(s)}\left(\boldsymbol\pi;\boldsymbol{v};\boldsymbol\theta\right)
&:=&
\sum_{l=1}^{K_{2}}\left(\frac{\pi_k}{\beta_{lk}}\right)^{\frac1{\varrho_l-1}}
\left[\sum_{j=1}^{K_1}\left(\frac{\pi_j^{\varrho_l}}{\beta_{lj}}\right)^{\frac1{\varrho_l-1}}
\right]^{-\frac1{\varrho_l}}
\left(v_l^{\frac{\varrho_l}{\tau_l}}-\beta_{l0}^{\varrho_l}\right)^{\frac1{\varrho_l}}
\end{eqnarray}
For input levels $\mathbf{x}$ and outputs $\mathbf{y}$, we constrain $v_k^{(0)}\equiv x_k$ for $k=1,\dots,K$ and $v_m^{(S)}\equiv y_m$ for $m=1,\dots,M$. If factor prices for inputs are observed, we also set $\pi_k^{(0)}\equiv p_k$, otherwise we treat prices as latent as well. In particular, we choose good 1 as the num\'eraire, $\pi_1^{(0)}=1$, so that for an interior solution to the cost minimization problem at node $1$ in layer $s=1$, the first-order conditions for optimal input choices require
\[\pi_k^{(0)}= \left(\frac{\beta_{1k}^{(1)}}{\beta_{11}^{(1)}}\right)^{\varrho_1^{(1)}}\left(\frac{v_{k1}^{(1)}}{v_{11}^{(1)}}\right)^{\varrho_1^{(1)}-1}\\
\nonumber=:\phi_{k}^{(0)}(\mathbf{v}^{(1)};\boldsymbol\theta_k^{(1)})\]
for $k=2,\dots,K$. Intermediate good 1 in layer 1 was chosen arbitrarily, so if only a subset of inputs are employed in nonzero quantities to produce that good, implied factor prices for the $K$ initial inputs can be inferred from the first-order conditions at other intermediate production nodes in layer $1$. Additional normalizations may be required if nodes in that layer only use non-overlapping subsets of the initial inputs.

To summarize, the production system is characterized by hidden states $h_k^{(s)}:=(v_k^{(s)},\pi_k^{(s)})$, where at the first stage $\pi_k^{(0)} = p_k$ and $v_k^{(0)}=x_k$ for $k=1,\dots,K_0\equiv K$, and at the last stage, $v_k^{(S)}=y_k$ for $k=1,\dots,K_{S}\equiv M$. Latent prices $\pi_k^{(s)}$ are determined recursively by iterating (\ref{prod_update_price}) forward from $s-1$ to $s$ starting at $s=1$, and quantities $v_k^{(s)}$ by iterating (\ref{prod_update_input}) backward from $s+1$ to $s$, starting at $s=S$.


\subsection{Data and Unobservables}

We consider training the network based on a sample of $n$ observed combinations $\mathbf{z}_i=(\mathbf{y}_i,\mathbf{x}_i')'$ of inputs and outputs. We allow for the possibility of measurement error in output levels as well as hidden inputs which we treat as stochastic. We therefore treat the technology $\mathcal{Y}^*$ as a random closed set and seek to approximate its (selection) expectation $\mathcal{Y}_0:=\mathbb{E}[\mathcal{Y}^*]$.\footnote{Following \cite{Mol05} the selection expectation of the random set $\mathcal{Y}^*:\Omega\rightarrow \mathcal{C}$ for some probability space $\Omega$ and the closed sets $\mathcal{C}$ in $\mathbb{R}^{K+M}$ is the closure of the set of all expectations of integrable selections, $\upsilon(\omega)\in\mathcal{Y}(\omega)$ for $\omega\in\Omega$.} It is known that when the distribution of $\mathcal{Y}^*$ is non-atomic, $\mathcal{Y}_0$ is a convex subset of $\mathbb{R}^{K+M}$ (see Theorem 1.15 in chapter 2 of \cite{Mol05}).

We furthermore assume that production is efficient in that any combinations of inputs (including unobserved inputs) and correctly measured outputs correspond to points on the efficient frontier of $\mathcal{Y}^*(\omega)$. We do not explicitly model the choice of a particular output combination on the efficient frontier in the multiple output case. Rather we assume that production and demand are separable, and at least in principle a model for demand could be estimated separately. We then match observed combinations of quantities to points on the frontier predicted by the ANN. Training loss is given by $\ell(\mathbf{z}_i,\boldsymbol\theta):=\sum_{m=1}^M(y_{mi} - v_{mi}^{(S)}(\boldsymbol\theta))^2$ where $v_{mi}^{(S)}(\boldsymbol\theta)$ is the activation of the $m$th neuron in the top layer given inputs $x_{1i},\dots,x_{Ki}$.


More generally, the $k$th observable covariate $X_{ki}$ could interpreted either as ``price"/cost shifter or the  quantity of an input for production, and matched accordingly to either the price $\pi_{ki}^{(0)}$ or the quantity $v_{ki}^{(0)}$ in the initial layer. An implementation of this neural network could also incorporate additional unobserved inputs at intermediate neurons, with one or multiple independent random draws from some continuous distribution with non-negative support. This would also define smooth likelihoods/energy functions for the intermediate stages for an implementation as a Deep Boltzmann machine. We are not aware of a ``conjugate" distribution for such an unobserved input that would yield closed-form solutions for factor demand or price equations at intermediate stages, but such an approach would in general have to rely on simulation by generating several replications for each unit, where data could also be permuted at random to reflect other invariance restrictions. While implementation is less straightforward, such an approach would at least conceptually be analogous to convolutional neural networks commonly used for image and video processing.

\section{Nested Discrete Decisions}

\label{sec:GEV_model}

Models for nested discrete decisions - binary or multinomial - have long been used to model individual choice behavior, see \cite{BAk73}, \cite{McF74}, and \cite{McF78}. \cite{HSW89} established that multi-layer feedforward networks with a ``squashing" activation function - a nondecreasing function mapping its scalar argument to the unit interval - can approximate any Borel measurable function on compacta. Our focus is on a flexible framework for multinomial discrete decisions, generalizing \cite{BAk73}'s nested Logit model to approximate all members of \cite{McF78}'s Generalized Extreme Value (GEV) class.


There are many examples when a single nesting structure may be too restrictive to model realistic cross-substitution patterns among a population of agents, especially when alternatives are complex or bundles of more elementary choices. For example when choosing a travel destination, Barcelona may appear to be a close substitute for Lisbon or Istanbul if the purpose of the trip is to combine a city trip with a beach vacation, however it may be a closer substitute to Amsterdam or Prague as a destination for party travel. \cite{Vov97} also motivates overlapping nests in a model of transportation choice when commuters may combine various modes of transportation for different legs (``trunk", ``egress") of the trip. Categorical outcomes may also result from choices among a richer set of heterogeneous alternatives - for example labor force status is determined by the decision whether to accept a particular job offer.

Standard random utility models (RUM) for discrete choice can be made more flexible to accommodate these richer substitution patterns, either by allowing for dependence among taste shocks, or modeling decisions sequentially via a richer latent hierarchy of nests of alternatives. Our modeling approach follows the second route but we also show that within \cite{McF78}'s Generalized Extreme Value (GEV) framework, the two are equivalent in terms of the distributions that a fully flexible model can generate.

\subsection{Nested Discrete Choice Model}

We consider a model of multinomial choice $Y$ among $M$ options $\{y_1,\dots,y_M\}$ given agent and alternative specific attributes $\mathbf{x}$. The object of interest are the conditional choice probabilities
\[\mu_0(y_m,\mathbf{x}):=\mathbb{P}(Y=y_m|\mathbf{X}=\mathbf{x}),\hspace{0.5cm}m=1,\dots,M\]
which are assumed to result from a random utility model (RUM) with a flexible dependence structure among taste shocks. Specifically, we let
\[U_{im}:=U_{im}^* + \varepsilon_{im},\hspace{0.3cm}m=1,\dots,M\]
with systematic parts $U_{im}^*:=U^*(\mathbf{x}_{im})$ that are nonstochastic functions of agent/alternative specific characteristics, and idiosyncratic taste shocks $\varepsilon_{im}$ that are independent of $\mathbf{x}_{im}$ with a common marginal distribution $G(\varepsilon)$. The systematic part $U_{im}^*$ can itself be specified flexibly, e.g. as an additively separable function of elements of $\mathbf{x}_{im}$.

By Sklar's theorem, the joint distribution of taste shocks for the $M$ alternatives can be written as
\[ G(\varepsilon_{1},\dots,\varepsilon_{M}) = C(G(\varepsilon_1),\dots,G(\varepsilon_{M}))\]
where the copula $C:[0,1]^{M}\rightarrow[0,1]$ is a $M$-non-decreasing function that is onto, where $C(0,\dots,0)=0$ and $C(1,\dots,u_m,\dots,1)=u_m$ for each $k=1,\dots,M$. Taking logs of joint and marginal c.d.f.s, we can rewrite the copula as
\[G(\varepsilon_{1},\dots,\varepsilon_{M}) = \exp\left\{-F_0\left(-\log G(\varepsilon_1),\dots,-\log G(\varepsilon_{M})\right)\right\}\]
where we refer to the mapping $F_0:\mathbb{R}_+^{M}\rightarrow\mathbb{R}_+$, $F_0:(w_1,\dots,w_{M})\mapsto F_0(w_1,\dots,w_{M})$ as the
\emph{generating function} associated with the copula $C(\cdot)$.


For the case when the marginal distribution $G(\varepsilon)=\exp\{-e^{-\varepsilon}\}$ is extreme-value type I, Theorem 1 in \cite{McF78} shows that any conditional choice probabilities of the form
\begin{equation}\label{mcf_gev_ccp}\mu_0(y_m,\mathbf{x}) = \frac{e^{U_{m}^*(\mathbf{x})}\frac{\partial}{\partial w_m}F_0\left(e^{U_{1}^*(\mathbf{x})},\dots,e^{U_{M}^*(\mathbf{x})}\right)}
{F_0\left(e^{U_{1}^*(\mathbf{x})},\dots,e^{U_{M}^*(\mathbf{x})}\right)}\end{equation}
can be generated by such a RUM under additional regularity conditions on the generating function $F_0(\cdot)$. Specifically, he assumes the following

\begin{ass}\label{GEV_generating_ass}\textbf{(GEV Generating Function)}
$F_0(\mathbf{w})$ is linearly homogeneous and $M$ times continuously differentiable, where $(-1)^{p+1}\frac{\partial^p}{\partial w_{m_1}\dots\partial w_{m_p}}F_0(\mathbf{w})\geq0$ for all $p=1,\dots,M$ and any distinct $m_1,\dots,m_p\in\{1,\dots,M\}$. In addition we assume that the same sign restriction on the cross-partial derivatives also holds with $m_1=m_2$ and $p\leq3$.
\end{ass}


The alternating sign condition on partial derivatives is satisfied by any function $F_0(\cdot)$ that is $M$ times differentiable and $M$-nondecreasing (see \cite{Nel06}, p.43 for a definition). Differentiability of degree greater than 1 is not needed for our results, whereas $M$-monotonicity is needed to ensure that the generating function defines a proper copula for the joint distribution of taste shocks under a random-utility interpretation of the conditional choice probabilities. In particular, these restrictions ensure that the mapping
\[C(u_1,\dots,u_{M})=\exp\left\{-F_0(-\log u_1,\dots,-\log u_M)\right\}\] of marginal ranks $u_1,\dots,u_M$ to $[0,1]$ is $M$-increasing and therefore yields a well-defined joint distribution function. However, these restrictions do not guarantee that $C(u_1,\dots,u_{M})$ satisfies the boundary conditions $C(1,\dots,u_m,\dots,1)=u_m$, so the function need not be a proper copula. Rather, permissible generating functions under Assumption \ref{GEV_generating_ass} may produce joint distributions with marginals that are different from the extreme value type-I distribution. We discuss the possibility of allowing for flexible, smooth specifications of the marginal distributions $G_1(\varepsilon),\dots,G_M(\varepsilon)$ in Appendix \ref{sec:GEV_flex_marg_app}.

We now propose a strategy to approximate the generating function flexibly using a model of multiple stages of discrete decisions. Following \cite{Vov97} and \cite{WKo01} we assume a nesting structure with multiple layers where alternatives can be grouped according to various, not necessarily exclusive aspects, rather we allow for each alternative to belong to multiple nests at a time. Overlapping nests may occur in settings when a heterogeneous population of agents with different nesting structures - in an example by \cite{Vov97} on the choice of transportation modes, an agent may assign a ``Park and Ride" commute either to a ``private" or ``public transit" nest, depending on which leg of the commute is more prominent.

We consider nested generating functions $F_k^{(s)}:\mathbb{R}_+^{K_{s-1}}\rightarrow\mathbb{R}_+$, where
\begin{eqnarray}
\nonumber F_k^{(s)}(w_1,\dots,w_{K_{s+1}})&\equiv&F_k^{(s)}(w_1,\dots,w_{K_{s+1}};\boldsymbol\theta_k^{(s)})\\
\label{CNL_generating_fct}&=&\left(\sum_{l=1}^{K_{s+1}}\left(\beta_{kl}^{(s)}w_l\right)^{\varrho_k^{(s)}}\right)^{\frac{1}{\varrho_k^{(s)}}}
\end{eqnarray}
where $\boldsymbol\theta_k^{(s)}:=(\beta_{k1}^{(s)},\dots,\beta_{kK_{s+1}}^{(s)},\varrho_k^{(s)})'$. The nesting coefficients $\varrho_k^{(s)}\geq1$ parameterize dependence of taste shocks within nests, where $\varrho_k^{(s)}=1$ corresponds to the case in which shocks are independent across nests, and $\varrho_k^{(s)}=\infty$ to the case of perfect rank dependence. The weights $\beta_{kl}^{(s)}$ can be interpreted as allocation parameters, measuring the relevance of nest $k$ for choosing alternative $l$. We then construct a generating function $F_{\mathbf{K}}^S(w_1,\dots,w_{M})$ recursively, where
\begin{eqnarray}
\nonumber w_k^{(S)}&=&w_k,\hspace{1.5cm}k=1,\dots,M\\
\label{CNL_recursion} w_k^{(s)}&=&F_k^{(s)}(w_1^{(s+1)},\dots,w_{K_{s+1}}^{(s+1)})
\end{eqnarray}
We can interpret this structure as a cross-nested Logit model with $S$ layers, where in each layer $s$ the agent chooses among $K_s$ latent nests $k_s=1,\dots,K_s$, and we also assume that the $0$th (bottom) layer consists of a single root node, i.e. $K_0=1$. The resulting nesting tree is a weighted directed, acyclic graph where a terminal node may be reached through various paths.

The conditional choice probabilities $\mu(y_m,\mathbf{x})$ are then approximated by
\begin{equation}\label{GEV_mu_KS}\mu_{\mathbf{K}}^S(y_m,\mathbf{x}):=\frac{e^{U_{m}^*(\mathbf{x})}\frac{\partial}{\partial w_m}F_{\mathbf{K}}^S\left(e^{U_{1}^*(\mathbf{x})},\dots,e^{U_{M}^*(\mathbf{x})}\right)}
{F_{\mathbf{K}}^S\left(e^{U_{1}^*(\mathbf{x})},\dots,e^{U_{M}^*(\mathbf{x})}\right)}\end{equation}
in analogy to (\ref{mcf_gev_ccp}). We allow for the possibility that $F_{\mathbf{K}}^S(\mathbf{w})$ is only directionally differentiable, in which case the partial derivative is taken to be the derivative from the right,
$\frac{\partial}{\partial w_m}F_{\mathbf{K}}^S(w_1,\dots,w_m,\dots,w_M):=
\lim_{t\downarrow0}\frac{F_{\mathbf{K}}^S(w_1,\dots,w_m+t,\dots,w_M)-F_{\mathbf{K}}^S(w_1,\dots,w_m,\dots,w_M)}{t}$.


We next show that with a sufficient number of latent nests, this model can in fact approximate the conditional choice probabilities resulting from the random utility model satisfying Assumption \ref{GEV_generating_ass} arbitrarily well.


\begin{prp}\label{CNL_nested_universal_prp}\textbf{(Universal Approximator)} Suppose that $F_0:\mathbb{R}_+^{M}\rightarrow\mathbb{R}_+$ satisfies Assumption \ref{GEV_generating_ass}. Then for any $\delta>0$ and compact rectangular set $\mathcal{C}\subset\textnormal{int} \mathbb{R}_{+}^M$, there exists an approximating function $F_{\mathbf{K}}^S$ constructed via (\ref{CNL_recursion}) with stage-wise generating functions of the form (\ref{CNL_generating_fct}) with $S=3$ such that
\[\sup_{\mathbf{x}\in\mathcal{C}}\left|\mu_{\mathbf{K}}^S(y_m,\mathbf{x})-\mu_0(y_m,\mathbf{x})\right|\leq\delta,\hspace{0.5cm}m=1,\dots,M\]
where the number of free parameters is $MK_2$, and
\[K_2 = c(R,M)\delta^{-\frac{M}{2}}\]
for a constant $c(R,M)$ that only depends on the dimension $M$ and diameter $R$ of $\mathcal{C}$.
\end{prp}

See the appendix for a proof. For the restriction of the approximation to arguments $w_1,\dots,w_M$ in the interior of the positive orthant, notice that (\ref{mcf_gev_ccp}) evaluates the generating function and its first partial derivatives only at arguments $w_m=\exp\{U_m^*(\mathbf{x})\}$ so that we can restrict our attention to compact subsets of $\textnormal{int}(\mathbb{R}_{+}^M)$ as long as $U_1^*(\mathbf{x}),\dots,U_{M}^*(\mathbf{x})$ are bounded functions of $\mathbf{x}$.

By (\ref{mcf_gev_ccp}) the conditional choice probabilities can be expressed as continuous functions of $F_0(w_1,\dots,w_M)$ and its first partial derivatives. Our proof is constructive in that we propose a four-layer neural network which approximates the subgraph of $F_0(w_1,\dots,w_M)$ on $\mathcal{C}$ with vertices corresponding to the $K_2$ neurons in the second layer, where the free weights determine the location of those support points, and the remaining layers generate the surface of that polytope using parameters which depend on $F(\mathbf{w})$ only through those weights from the second layer. We then use known results on approximations of convex sets by polyhedra to bound the resulting error regarding the function $F_0(w_1,\dots,w_M)$ and its first partial derivatives.


It is also important to note that for this approximation, nest-specific weights $\beta_{kl}^{(s)}$ are non-negative, and nesting coefficients satisfy $\varrho_k^{(s)}\geq 1$, so that the approximating model is consistent with random utility maximization. One interpretation of the conventional nested Logit model represents the taste shifters for the alternatives in the final layer as a sum of independent alternative- and nest-specific random utility shocks (see \cite{BLe85} and \cite{Gal21}). We can therefore think of the deep network approximating the ``true" error distribution with a convolution of independent shocks some of which are shared by subsets of the $M$ alternatives.

\subsection{Implementation}

To nest this model into the recursive framework in (\ref{forward_iteration}) and (\ref{backward_iteration}), we parameterize conditional choice probabilities from maximizing behavior in terms of inclusive values. Specifically we define the inclusive value for nest $k$ in the $s$th layer recursively via
\begin{eqnarray}
\nonumber v_k^{(S)}&:=&e^{U_k^*}\\
\label{nest_v_recursion}
v_k^{(s)}&:=&\psi_k^{(s)}(v_1^{(s+1)},\dots,v_{K_{s+1}}^{(s+1)};\boldsymbol\theta_k^{(s)})
\end{eqnarray}
where
\begin{equation}
\label{nest_inclusive_val_psi} \psi_k^{(s)}(\mathbf{v}^{(s+1)};\boldsymbol\theta_k^{(s)}):=
\log\left(\left[\sum_{l=1}^{K_{s+1}}\left(\beta_{kl}^{(s)}e^{v_l^{(s+1)}}\right)^{\varrho_k^{(s)}}\right]^{1/\varrho_k^{(s)}}\right)
\end{equation}
Note that we can take limits
\[\lim_{\varrho\rightarrow\infty} \left[\sum_{l=1}^{K}\left(\beta_{l}e^{v_l}\right)^{\varrho}\right]^{1/\varrho}
=\max\left\{\beta_le^{v_1},\dots,\beta_Ke^{v_K}\right\}\]
so that for large values of $\varrho_{k}^{(s)}$, we can interpret $\psi_k^{(s)}$ in (\ref{nest_inclusive_val_psi}) as a softmax mapping from $e^{v_1^{(s+1)}},\dots, e^{v_{K_{s+1}}^{(s+1)}}$ to $e^{v_k^{(s)}}$.

We can next characterize the conditional choice probabilities from this random utility model recursively in terms of these inclusive values. Specifically, we define the probabilities $\pi_k^{(s)}$ of reaching the $k$th nest in the $s$th layer via the recursion
\begin{eqnarray}
\nonumber\pi_1^{(0)}&:=&1\\
\label{nest_pi_recursion} \pi_k^{(s)}&:=&\phi_k^{(s)}(\mathbf{v}^{(s)},\boldsymbol{\pi}^{(s-1)};\boldsymbol\theta_k^{(s)})
\end{eqnarray}
initialized at the root node. Lemma \ref{ccp_phi_lem} below shows that this mapping is given by
\begin{equation}
\label{nest_ccp_phi}\phi_k^{(s)}(\mathbf{v}^{(s)},\boldsymbol{\pi}^{(s-1)};\boldsymbol\theta_k^{(s)})
:=\sum_{l=1}^{K_{s-1}} \frac{\left(\beta_{kl}^{(s)}e^{v_l^{(s-1)}}\right)^{\varrho_k^{(s)}}}
{\sum_{m=1}^{K_{s-1}}\left(\beta_{km}^{(s)}e^{v_m^{(s-1)}}\right)^{\varrho_k^{(s)}}}\pi_l^{(s-1)}
\end{equation}

\begin{lem}\label{ccp_phi_lem} Consider the nested model in (\ref{CNL_recursion}). Then the conditional choice probability for alternative $y_m$ is given by $\pi_m^{(S)}$, with intermediate states $h_k^{(s)}:=(\pi_k^{(s)},v_k^{(s)})$ characterized recursively by the mappings (\ref{nest_inclusive_val_psi}) and (\ref{nest_ccp_phi}). 
\end{lem}

See the appendix for a proof. It is interesting to note that the recursion defining inclusive values and conditional choice probabilities is triangular in that the mapping iterating inclusive values (\ref{nest_inclusive_val_psi}) does not take conditional choice probabilities $\pi_k^{(s)}$ as an argument, so that all hidden states can be calculated with a single backward pass followed by a forward pass of the recursions $\psi_k^{(s)}(\cdot)$ and $\phi_k^{(s)}(\cdot)$, respectively.

The main object of interest in the nested GEV model is the conditional choice probability
\[\mu_0(y,\mathbf{x}):=\mathbb{P}(Y=y|\mathbf{X}=\mathbf{x})\]
Given the trained network, the model prediction for $\mu_0(y,\mathbf{x})$ is then given by the activations
\[\mu_{\mathbf{K}}^S(y,\mathbf{x};\boldsymbol\theta):=\left\{\begin{array}{lcl}\pi_{m}^{(S)}(\mathbf{x};\boldsymbol\theta)&\hspace{0.5cm}&\textnormal{if }y=y_m\textnormal{ for some }m=1,\dots,M\\
0&&\textnormal{if }y\notin\{y_1,\dots,y_M\}\end{array}\right.\]
In order to train the network, we can therefore use the activations $\pi_1^{(S)}(\mathbf{x},\boldsymbol\theta),\dots,\pi_M^{(S)}(\mathbf{x},\boldsymbol\theta)$ in the top layer in a regression layer with loss $\ell(\mathbf{z},\boldsymbol\theta):=-\sum_{m=1}^M\dum\{y=y_m\}\log\pi_{m}^{(S)}(\mathbf{x},\boldsymbol\theta)$ and deeper layers specified according to (\ref{nest_v_recursion}) and (\ref{nest_pi_recursion}).

\begin{rem}\textbf{(Aggregation over Heterogeneous Alternatives)} So far we have treated the number of alternatives in the multinomial choice problem as given. We next extend this framework to a problem where the observable alternatives $k=0,\dots,K_S$ aggregate a larger number of heterogeneous alternatives. For example, in residential choice, we may only observe a coarse partition of the set of available alternatives, with the agent choosing among the alternatives that are optimal within each category.

Specifically, we could consider a fixed set of observable \emph{categories} $\{1,\dots,M\}$ which are a partition of a set of primitive alternatives $k=1,\dots,K_S$, where $m:\{1,\dots,K_S\}\rightarrow\{1,\dots,M\}$ assigns each alternative to one of the $M$ categories. Random utility for alternative $k$ is given by
\[U_{ik}:=x_i'\gamma_{k}^{(S)} + h^{(S)}(z_{m(k)}) + \varepsilon_{ik},\hspace{0.3cm}k=1,\dots,K_S\]
where coefficients $\beta_k$ may vary within each category of alternatives, and $h(z)$ is a smooth function of alternative-specific characteristics. We will also generally assume that $h(z)$ is additively separable in the scalar components of $z$. We interpret the alternative-specific coefficients $\beta_{k}^{(S)}$ is as unobserved attributes or ``amenities" of each primitive alternative that vary in relevance to agents with different attributes $x_i$.

The conditional probability of choosing an alternative in category $m$ is then given by
\[\mathbb{P}\left(\max_{m(k)=m}U_{ik}\geq\max_{k}U_{ik}\right) = \sum_{k:m(k)=m}
\mathbb{P}\left(U_{ik}\geq\max_{k}U_{ik}\right)\]
Allowing for $K_S$ to grow large would provide additional flexibility in approximating conditional choice probabilities among the $J$ categories.

\end{rem}

\section{Separability and Sparsity Restrictions}

%
%

Generally speaking, whether or not there is an advantage in approximating the relationship between inputs and outputs using a nested model which mimics this structure depends on whether there are any meaningful constraints on the nested mappings $\phi_k(\cdot)$ and $\psi_k(\cdot)$ in the generative model. Specifically, we consider qualitative restrictions on the intermediate transformation functions $\left\{F_k^{(s)}(w_1,\dots,w_{K_{s-1}})\right\}_{k,s}$ either in production or the construction of the GEV generating function, which are the main economic primitives of either problem.

Our general approach already makes use of global shape restrictions - monotonicity, convexity, homogeneity where applicable - on certain economic primitives. Generally speaking, qualitative restrictions on components of a structural model do not necessarily translate into analogous restrictions that could be imposed on the reduced form in a straightforward manner. However, for the flexible approximating models proposed in this paper we show that these shape constraints can be incorporated quite naturally into estimation as sign restrictions on model parameters and yield a substantial reduction of its statistical complexity.

In this section we consider two additional qualitative/nonparametric restrictions that formalize how economic primitives at individual stages of this model may be more ``fundamental" than their composition.

\begin{dfn}\textbf{(Sparsity)}\label{sparse_dfn} (a) We say the \emph{function} $F^{(s)}:\mathbb{R}^{K_{s-1}}\rightarrow\mathbb{R}$ is $d^*$-sparse if it can be written as a function of at most $d^*$ of its arguments, that is there exist $k_1,\dots,k_{d^*}$ and $\tilde{F}^{(s)}:\mathbb{R}^{d^*}\rightarrow\mathbb{R}$ such that
\[F^{(s)}(w_1,\dots,w_{K_{s-1}})\equiv\tilde{F}^{(s)}(w_{k_1},\dots,w_{k_{d^*}})\]
for all $w_1,\dots,w_{K_{s-1}}$. (b) A \emph{MLP} of $S$ layers is said to be $d^*$-sparse if each hidden layer consists of at most $d^*$ neurons, each of which are a $d^*$-sparse function $F_k^{(s)}(\mathbf{w}^{(s-1)})$ of inputs from the preceding layer.
\end{dfn}

In models of production, sparsity reflects that intermediate technologies may only use some of the available inputs for production. For nested models of discrete choice, sparsity restricts the cross-nesting structure by allowing any nest to have at most $d^*$ successors in the following nesting layer. Sparsity is assumed for the results in \cite{MPo16} and \cite{BKo19} with $d^*=2$, and general $d^*\leq K$, respectively.

\begin{dfn}\textbf{(Separability)}\label{separable_dfn} (a) We say that the \emph{function} $F^{(s)}:\mathbb{R}^{K_{s-1}}\rightarrow\mathbb{R}$ is separable in its arguments if there exist $K_{s-1}$ functions $g_k^{(s)}:\mathbb{R}\rightarrow\mathbb{R}$ and a transformation $f^{(s)}:\mathbb{R}\rightarrow\mathbb{R}$ such that
\[F^{(s)}(w_1,\dots,w_{K_{s-1}})\equiv f^{(s)}(g_1^{(s)}(w_1)+\dots +g_{K_{s-1}}^{(s)}(w_{K_{s-1}}))\]
for all $w_1,\dots,w_{K_{s-1}}$, where $f^{(s)}(\cdot),g_k^{(s)}(\cdot)$ are Lipschitz continuous with Lipschitz constant $\lambda<\infty$. (b) A \emph{MLP} of $S$ layers is said to be separable if each hidden layer consists of at most $K$ neurons, each of which are a separable function $F_k^{(s)}(\mathbf{w}^{(s-1)})$ of inputs from the preceding layer.
\end{dfn}

Separability is satisfied by index models $F(\mathbf{v}'\boldsymbol\beta)$ for arbitrary link functions $F(\cdot)$ and the constant-elasticity of substitution (CES) family of production functions, including the cases of perfect complements and perfect substitutes. When the functions $g_1^{(s)}(\cdot),\dots,g_{K_{s-1}}^{(s)}(\cdot)$ are the identity, this corresponds to the nested nonparametric regression model analyzed in \cite{HMa07}. In principle, this definition can be extended to include models satisfying separability among partitions of $w_1,\dots,w_{K_{s-1}}$ into subsets of multiple variables.

One key observation is that neither of these properties of link functions $F_k^{(s+1)},F_l^{(s)},\dots,F_{K_{s}}^{(s)}$ is generally inherited by compositions
\[F_k^{(s,s+1)}(w_1,\dots,w_{K_{s-1}}):=F_k^{(s+1)}(F_1^{(s)}(w_1,\dots,w_{K_{s-1}}),\dots,F_{K_{s}}^{(s)}(w_1,\dots,w_{K_{s-1}}))\]
Moreover, our main focus will be on cases in which the relevant intermediate state variables $v_k^{(s)},\pi_k^{(s)}$ are the result of optimization with an objective function $F_k^{(s)}(\cdot)$ satisfying sparsity or separability, whereas the activation functions $\psi_k^{(s)}(\cdot),\phi_k^{(s)}$ derived from these primitives need in general not exhibit either property. It is for these reasons that we aim to construct the approximating network to mirror the nested architecture of the data generating process when applicable.

We now propose network architectures which directly impose either sparsity or separability within components of an $S$-layer generative model with neurons of unspecified functional form. Within our framework, it is then possible to construct a common architecture which would nest all three cases - sparse, separable, and unrestricted - where either restriction can be imposed by setting weights for certain connections equal to zero. For simplicity, we state our results for estimation of production models, the case of discrete choice models is entirely analogous.

\subsection{Sparsity}

We first consider approximation of a model with stage technologies that are $d^*$-\emph{sparse} according to Definition \ref{sparse_dfn} with dimension $d^*\leq K$, i.e. a production or generating function of the form
\[F_{k}^{(s)}(w_1,\dots, w_{K_{s-1}})=f_k^{(s)}(w_{k_1},\dots w_{k_{d^*}}) \]
Following the construction in \cite{BKo19}, we propose the following approximating network: 
\begin{itemize}
\item We use a three-layer network $\hat{F}_k^{(s)}$ to approximate $F_k^{(s)}$ where the bottom layer takes $w_1,\dots, w_{K_{s-1}}$ as inputs and has $K_1=3d^*$ neurons.
\item The middle layer consists of $K_2=Q$ neurons (independent of $K$), and
\item the top layer consists of a single linear neuron with $\varrho_1^{(3)}=\tau_1^{(3)}=1$. All other parameters are left unrestricted.
\item Finally, we assemble the full MLP $F_{\mathbf{K},Q}^S$ consisting of $S$ layers of three-stage neurons $\hat{F}_k^{(s)}$, $k=1,\dots,d^*$ and $s=1,\dots,S$.
\end{itemize}

We can now state the main result regarding the rate of approximation for a sparse stage production technology given that network architecture:


\begin{prp}\label{sparse_appx_prp}\textbf{(Sparse Stage Technologies)} Suppose that the technology $\mathcal{Y}_0$ can be characterized by a $d^*$-sparse MLP with $S$ layers, where $F_k^{(s)}(\mathbf{w}^{(s-1)})$ are concave, nondecreasing and uniformly Lipschitz. Then for any $\delta>0$ and compact rectangular set $\mathcal{C}\subset\textnormal{int}\mathbb{R}_{+}^{K+M}$, there exists an approximating function
$F_{\mathbf{K},Q}^S$ constructed on a network of the type (SP-CES) with depth $3S$ such that
\[\sup_{(\mathbf{u},\mathbf{x})\in\mathcal{C}}\left|\mu_{\mathbf{K}}^S(\mathbf{u},\mathbf{x})-\mu_0(\mathbf{u},\mathbf{x})\right|\leq\delta\]
where
\[Q = c(R,S,d^*)\delta^{-\frac{d^*}{2}}\]
and $c(R,S,d^*)$ is a constant that only depends on the dimension $d^*$ and diameter $R$ of $\mathcal{C}$.
\end{prp}

See the appendix for a proof. Comparing this result to the rates in Proposition \ref{CES_prod_universal_prp}, all rates are in terms of the effective dimensionality $d^*$ under sparsity rather than $K$. An analogous result can be given for the discrete choice model under Assumption \ref{GEV_generating_ass} when the generating function is $d^*$-sparse.

\subsection{Separability}

We next consider separability restrictions in the production model. For models of discrete decisions, these restrictions can be imposed in an entirely analogous manner. In order to approximate an intermediate production technology that is \emph{separable} in the sense of Definition \ref{separable_dfn}, i.e. satisfying
\[F_k^{(s)}(w_1,\dots,w_{K_{s-1}}) = f_k^{(s)}(g_{k1}^{(s)}(w_1)+ \dots + g_{kK_{s-1}}^{(s)}(w_{K_{s-1}}))\]
we propose the following Four-layer CES (4L-CES) specification of the production network uses three layers of nested CES production functions to approximate $F_k^{(s)}$.
\begin{itemize}
\item The bottom layer consists of $Q$ neurons with $\tau_q^{(1)}=1$ and the remaining parameters $\boldsymbol\beta^{(1)},\boldsymbol\varrho^{(1)}$ unrestricted.
\item The second layer consists of a single neuron with $\tau^{(2)}=\varrho^{(2)}=1$ and the remaining parameters $\boldsymbol\beta^{(2)}$ varying freely.
\item The third layer consists of $Q$ neurons with $\tau_q^{(3)}=1$ and the remaining parameters $\boldsymbol\beta^{(3)},\boldsymbol\varrho^{(3)}$ unrestricted.
\item The fourth (top) layer is linear and consist of a single neuron with $\varrho^{(3)}=\tau^{(3)}=1$ and $\boldsymbol\beta^{(2)}$ varying freely. These four layers form a self-contained module which does not connect to those approximating other stage technologies in the same layer.
\item Finally, we assemble the full MLP consisting of $S$ layers of three-stage neurons $\hat{F}_k^{(s)}$, $k=1,\dots,d^*$.
\end{itemize}

Using this construction we can achieve the following approximation rates for separable stage technologies:

\begin{prp}\label{separable_appx_prp}\textbf{(Separable Stage Technologies)} Suppose that the technology $\mathcal{Y}_0$ can be characterized by a MLP with $S$ layers of at most $K$ neurons, where $F_k^{(s)}(\mathbf{w}^{(s-1)})$ are separable and uniformly Lipschitz, with concave and nondecreasing link functions $f_k^{(s)}(\cdot)$ and $g_{kl}^{(s)}(\cdot)$. Then for any $\delta>0$ and compact rectangular set $\mathcal{C}\subset\textnormal{int}\mathbb{R}_{+}^{K+M}$, there exists an approximating function
$F_{\mathbf{K},Q}^S$ constructed on a network of the type (4L-CES) with depth $4S$ such that
\[\sup_{(\mathbf{u},\mathbf{x})\in\mathcal{C}}\left|\mu_{\mathbf{K}}^S(\mathbf{u},\mathbf{x})-\mu_0(\mathbf{u},\mathbf{x})\right|\leq\delta\]
where
\[Q = c(R,S,K)\delta^{-\frac{1}{2}}\]
and $c(R,S,K)$ is a constant that only depends on the dimension $d^*$ and diameter $R$ of $\mathcal{C}$.
\end{prp}

See the appendix for a proof.


%

\section{Asymptotic Theory}

This section gives convergence rates for estimating the function $\mu_0(\mathbf{y},\mathbf{x})$, where we approximate the underlying model primitives - the feasible set $\mathcal{Y}_0$ or the generating function $F_0(\mathbf{w})$, respectively - with a multilayer perceptron. We consider the class $\mathcal{H}_{\mathbf{K}}^S$ of multilayer neural networks with $S$ layers, $\mathbf{K}=(K_1,\dots,K_S)$ hidden nodes, and $W$ free parameters (weights $\beta_k^{(s)}$ and coefficients $\varrho_k^{(s)}$). We denote the resulting approximation to the target function with $\mu_{\mathbf{K}}^S(\mathbf{y},\mathbf{x})$. Note that in implementing this approach we do not explicitly compute and report the full set $\mathcal{Y}_0$ or generating function $F_0(\mathbf{w})$, rather the trained network only evaluates whether any given point in the sample belongs to that set. Global shape restrictions on that set then only restrict complexity of ways of assigning points in and outside of that set.



Estimation of $\mu_0$ is based on a sample of $n$ observations, where we assume the following:

\begin{ass}\label{data_ass} (a) The sample $\mathbf{z}_1,\dots,\mathbf{z}_n$ consists of i.i.d. realizations of a random variable $\mathbf{Z}=(\mathbf{X},\mathbf{Y})$, where the p.d.f. of $X$ is bounded away from zero on its support $\mathcal{X}\equiv[-1,1]^K$. (b) For the production model we furthermore assume that $\mathcal{Y}_0(\mathbf{x})\subset[0,B]^M$  for all $\mathbf{x}\in\mathcal{X}$ and that $\mathbf{Z}$ has full support on the boundary of $\mathcal{Y}_0\cap\mathcal{X}\times\mathbb{R}^M$. (c) For the discrete choice model we also assume that random utilities satisfy $|U_m^*(\mathbf{x})|\leq B$ for some $B<\infty$ and all $m=1,\dots,M$.
\end{ass}

The assumption of a rectangular support for $X$ is primarily for analytical convenience and can be generalized to other compact subsets of $\mathbb{R}^K$ with nonempty interior. The condition that we observe input/output combinations over the entire efficient frontier of $\mathcal{Y}_0$ requires additional assumption on the mechanism for selection among multiple efficient input/output combinations, e.g. due to sufficient variation in input and output prices for a profit-maximizing producer. A more rigorous formulation in terms of economic primitives is beyond the scope of this paper and will be left for future research. If that condition fails, the researcher may still learn about a segment of the efficient frontier determining $\mu_0(\mathbf{u},\mathbf{x})$ for certain directions $\mathbf{u}$ and a restricted set of input combinations $\mathbf{x}$.


Our asymptotic results then concern the estimate of $\mu_0(\mathbf{y},\mathbf{x})$ based on that sample. Following \cite{FLM19}, suppose that we train the network according to a loss function $\ell(\mu,\mathbf{z})$ such that
\[\mu_0=\argmin_{\mu}\mathbb{E}[\ell(\mu,\mathbf{Z})].\]
In addition we assume that $\ell(\cdot,\cdot)$ is Lipschitz and that there exist $c_1,c_2>0$ such that
\[c_1\mathbb{E}\left[(\mu-\mu_0)^2\right]\leq\mathbb{E}\left[\ell(\mu,\mathbf{Z})\right]-\mathbb{E}\left[\ell(\mu_0,\mathbf{Z})\right]
\leq c_2\mathbb{E}\left[(\mu-\mu_0)^2\right]\]
After training the network $\mathcal{H}_{\mathbf{K}}^S$ we obtain the estimator
\begin{equation}\label{mu_hat_def}\hat{\mu}_n\in\argmin_{\mu\in\mathcal{H}_{\mathbf{K}}^S\\ \|\tau\|_{\infty}\leq 2B}\frac1n\sum_{i=1}^n\ell(\mu,\mathbf{z}_i)\end{equation}

\subsection{VC Dimension}

To characterize the asymptotic properties of $\hat{\mu}_n$ we start by stating a general VC bound of the approximating neural network which applies to general convex approximating architectures.

\begin{lem}\label{convex_VC_bound} For the class $\mathcal{H}_{\mathbf{K}}^S$ of networks with $W$ free parameters satisfying $\beta_{k}^{(s)}\geq0$ and $\varrho_k^{(s)}\leq1$ for all $k,s$, the VC dimension is bounded by
\[VCdim(\mathcal{H}_{\mathbf{K}}^S)= O\left(2W\log(2e)\right)\]
\end{lem}

See the appendix for a proof. The proof of this result applies the generic bound in Theorem 2 by \cite{KMa97}, noting that for any concave function all level and contour sets are connected, so that in the notation of their paper $B=1$. Furthermore, it is important to note that their results apply to Boolean combinations of ``atomic formulas" which are arbitrary, infinitely differentiable functions of the data and parameters. In particular, the theorem does not require the $\mu_{\mathbf{K}}^S(\mathbf{y},\mathbf{x})$ to be defined in terms of an acyclic (feedforward) multilinear perceptron.


Other restrictions, like sparsity or separability enter through the architecture of the neural network, where the constrained network requires only a smaller number $W$ of adjustable weights. Since Lemma \ref{convex_VC_bound} depends only on $W$ with no additional assumptions on network architecture, VC bounds for shape constrained networks can be obtained from that same result.

\begin{rem} It is useful to compare this VC bound to ``generic" VC bounds for a neural network with $S$ layers and $W$ adjustable weights without shape constraints: If the coefficients $\beta_{jl}^{(s)}$ are unrestricted, $1\leq\tau_l^{(s)}\leq k$ are integer, and $\varrho_l^{(s)}\in\{-\infty,1,\infty\}$, the resulting network corresponds to one with a piecewise polynomial activation function of degree $k$ or less. It then follows from Theorem 1 in \cite{BMM98} that the VC dimension of the resulting network is of the order no greater than
\[VCdim(\mathcal{H})= O\left(WS\log W + WS^2\right)\]
For non-integer values of $\tau_l^{(s)}$ the resulting activation functions are no longer polynomial. Rather, we can use the fact that power functions can be embedded into Pfaffian chains of length 1, so that from section 4.2 in \cite{KMa97} it follows that the VC dimension grows at a rate no faster than
\[VCdim(\mathcal{H})= O\left(W^2m^2\right)\]
\cite{KMa97} also suggest that this rate may not be sharp but conjecture that it may be imrpoved to match that for the piecewise polynomial case even for general Pfaffian activation functions.
\end{rem}

Most notably the bound in Lemma \ref{convex_VC_bound} implies no penalty for network depth besides through the number of adjustable parameters, whereas all of the generic results (allowing for negative coefficients) do. The main reason for this is that without convexity or monotonicity, lower contour sets of the functions a neural network may generate consist of multiple connected components, where the number of connected components increases with network depth.

\subsection{Convergence Rate}
Given the previous results regarding the rate of approximation and VC dimension of these networks, we can adapt the proofs for the main results in \cite{FLM19} to obtain the asymptotic rate for estimation of $\mu_0(\cdot)$. Our approximation results in Propositions \ref{CES_prod_universal_prp} and \ref{CNL_nested_universal_prp} and Lemma \ref{convex_VC_bound} will simply take the place of Lemmas 6 and 7 in their argument, which is otherwise not specific to the case of ReLU feedforward networks.

\begin{thm}\label{conv_rate_thm}\textbf{Convergence Rate} Suppose that Assumption \ref{data_ass} (a) holds, and let $\hat{\mu}_n\in\mathcal{H}_{K_1}^S$ be the estimator defined in (\ref{mu_hat_def}). If in addition, the model satisfies Assumptions \ref{technology_ass} and \ref{data_ass} (b), there exists a constant $C>0$, independent of $n$ such that for $d=K+M-1$, $S=2$ and $K_1\asymp\left(\frac{n}{\log n}\right)^{\frac{d}{4+d}}$ we have that for a compact rectangular set $\mathcal{C}\subset\mathbb{R}^d$,
\[\|\hat{\mu}_n-\mu_0\|_{\mathcal{C},2}^2\leq C\left(\frac{\log n}n\right)^{\frac{4}{4+d}}\]
with probability approaching 1. If the model satisfies Assumptions \ref{GEV_generating_ass} and \ref{data_ass} (c), then the analogous conclusion holds with $d=M$ and $S=3$.
\end{thm}

We state the result for the smallest number of hidden layers required for the approximation in \ref{CES_prod_universal_prp} and \ref{CNL_nested_universal_prp}, noticing that neither result suggests an added benefit to depth for the case in the absence of additional shape restrictions. However, that conclusion changes once we assume that stage technologies are sparse or separable, and we derive convergence rates for that case separately below. Note also that the rate for $\hat{\mu}_n$ matches the minimax risk bound for nonparametric estimation of convex functions in Theorem 2.4 by \cite{HWe16}, up to logarithmic terms. Interestingly, they also point out that the default nonparametric estimator, bounded least squares, fails to achieve that bound for $d>4$.

As discussed in the context of Theorem 3 in \cite{FLM19}, the approximation bounds underlying this result can also be strengthened in order to establish asymptotic normality and variance estimation for certain functionals of $\mu_0(\cdot)$. Since apart from the separate derivation of the VC dimension of $\mathcal{H}_{\mathbf{K}}^S$ and its approximating properties the argument is completely analogous to their case we do not prove the analogous conclusions separately for the present framework.

In order to appreciate the gains from imposing additional shape restrictions, we next state the convergence rates under sparsity and separability restrictions for the stage functions in an $S$-layer model with neurons of unknown functional form.

\begin{cor}\textbf{Convergence Rate under Sparsity} Suppose that Assumption \ref{data_ass} holds, and let $\hat{\mu}_n\in\mathcal{H}_{Q}^{(3S)}$ be the estimator based on the network (SP-CES). If the model satisfies Assumption \ref{technology_ass} and is $d^*$-sparse with $d^*\leq K+M-1$, there exists a constant $C>0$, independent of $n$ such that for $S=2$ and $Q\asymp\left(\frac{n}{\log n}\right)^{\frac{d^*}{4+d^*}}$ we have that for a compact rectangular set $\mathcal{C}\subset\mathbb{R}^d$,
\[\|\hat{\mu}_n-\mu_0\|_2^2\leq C\left(\frac{\log n}n\right)^{\frac{4}{4+d^*}}\]
with probability approaching 1. If the model satisfies Assumption \ref{GEV_generating_ass} and is $d^*$-sparse with $d^*\leq M$, then the analogous conclusion holds for $S=3$.
\end{cor}

This result follows immediately from the proof of Theorem \ref{conv_rate_thm}, noting that $Q$ can be chosen according to the approximation rate in Proposition \ref{sparse_appx_prp} and that the resulting approximating network has less than $3Sd^*Q$ free parameters.

\begin{cor}\textbf{Convergence Rate under Separability} Suppose that Assumption \ref{data_ass} holds, and let $\hat{\mu}_n\in\mathcal{H}_{Q}^{(4S)}$ be the estimator based on the network (4S-CES). If the model satisfies Assumption \ref{technology_ass} or Assumption \ref{GEV_generating_ass} and consists of separable stage functions, there exists a constant $C>0$, independent of $n$ such that for $S=2$ and $Q\asymp\left(\frac{n}{\log n}\right)^{\frac{1}{5}}$ we have that for a compact rectangular set $\mathcal{C}\subset\mathbb{R}^d$,
\[\|\hat{\mu}_n-\mu_0\|_2^2\leq C\left(\frac{\log n}n\right)^{\frac{4}{5}}\]
with probability approaching 1. If the model satisfies Assumption \ref{GEV_generating_ass}, then the analogous conclusion holds for $S=3$.
\end{cor}

As for the $d^*$-sparse case, this result follows again from the proof of Theorem \ref{conv_rate_thm}, where $Q$ can be chosen according to the approximation rate in Proposition \ref{separable_appx_prp}. The resulting approximating network has less than $(K+5)SQ$ free parameters, so that we obtain the conclusion by applying that rate to the generic VC bound in Lemma \ref{convex_VC_bound}.

\section{Conclusion}

This paper explores the use of artificial neural networks to approximate the reduced form of models of production and discrete decisions. For one we illustrate the expressive capacity of nested models of optimizing behavior, where in the absence of additional restrictions on the number of hidden units, such a model can generate any reduced form within a nonparametric class satisfying only broad shape constraints. Conversely that approximating property can be used for estimation, where nested models can be used to approximate an otherwise unconstrained reduced form. Here our results imply that the reduced form can be estimated at an asymptotic rate corresponding to the optimal nonparametric rate for estimating regression functions with bounded partial derivatives up to order two. Furthermore, monotonicity and convexity can be imposed as simple sign restrictions on model parameter.

One important benefit of using an approximating function class that mirrors a structural model for the data generating process is that additional sparsity or separability restrictions can be imposed directly on the network architecture, resulting in a faster rate of convergence for the estimator. This raises the question whether the network architecture can be made adaptive to the unknown structure of the underlying model, e.g. using $L^{1}$-penalization of weights, which will be left for future research.

\bibliographystyle{econometrica}
\bibliography{mybibnew}

\footnotesize
\appendix

\section{Training and Computation}

\label{sec:backprop_app}

Since some versions of the MLP proposed in this paper contain some states that iterate forward and others that iterate backwards, the gradient of the objective function cannot be evaluated directly using the backpropagation algorithm. We discuss an interpretation of the classical backpropagation algorithm as a recursive computation of the matrix inverse of the Jacobian for the system of fixed point equations characterizing all hidden and manifest states of the network. We then argue that the same approach continues to provide an arbitrarily good approximation to the gradient under a stability condition even when the network graph may contain cycles. This generalization of the the classical backpropagation algorithm was first proposed by \cite{Alm87} and \cite{Pin87}; \cite{LXF18} provided a more recent appraisal. A memory-efficient implementation of deep equilibrium models using root-finding algorithms was recently proposed by \cite{BKK19}.

In order to train the model, we use an algorithm that iterates between three steps:
\begin{itemize}
\item given parameters $\boldsymbol\theta_t$, we update the hidden states of the network $\boldsymbol\pi_t,\mathbf{v}_t$.
\item given states $\boldsymbol\pi_t,\mathbf{v}_t$, we compute the gradient of the loss function with respect to parameters $\boldsymbol\theta$, $\nabla_{\boldsymbol\theta}L(\boldsymbol\theta_t)$.
\item We update weights using a learning rule
\[\theta_{t+1} = \theta_t + \eta\nabla_{\boldsymbol\theta}L(\boldsymbol\theta_t)\]
given the learning rate $\eta$.
\end{itemize}

To compute the gradient with respect to weights $\boldsymbol\theta$, we propose a generalization of the back-propagation algorithm, which was first proposed for training neural networks by \cite{RHW86}. In multi-layer feedforward networks, back-propagation computes the partial derivatives of the training criterion with respect to all neuron weights using the chain rule. In typical implementations of deep networks, the number of weights is large, and standard numerical differentiation would require re-evaluating the network separately for each weight, whereas back-propagation can compute partial derivatives with respect to all weights simultaneously via a single pass through the network. Since the hidden units are only updated recursively, the first approach becomes prohibitively time intensive even for networks of only moderate depth.

Back-propagation exploits the recursive structure of the network, however it is not directly applicable to our problem since there are two sets of network states, one that is determined by iterating forward, the other by backward induction. There are some special cases (e.g. with bidirectional recurrent networks) in which those two sets of states do not interact and the backward propagation algorithm can be run simultaneously on two non-overlapping networks in opposing directions. However nested optimization models of the kind considered in this paper generally exhibit feedback loops where hidden states in multiple layers are determined simultaneously rather than recursively, and the Jacobian matrix of partial derivatives of activation functions and hidden states cannot be made triangular.

We argue that feedback cycles are not an unsurmountable obstacle to the application of a back-propagation algorithm, but rather that an iterative algorithm of this type can be used to approximate the gradient of the criterion function arbitrarily well after finitely many steps. Specifically, the recursive application of the chain rule in the classical back-propagation algorithm can be interpreted as the von-Neumann series approximating a matrix inverse, where in the special case of the Jacobian for a feedforward network, that approximation becomes exact after $S$ steps.

Specifically, we can represent the network in (\ref{forward_iteration}) and (\ref{backward_iteration}) as a system of equations
\[\left[\begin{array}{c}\boldsymbol\pi\\ \boldsymbol{v}\end{array}\right]=\left[\begin{array}{c}\boldsymbol{\phi(\pi,v;\theta)}\\ \boldsymbol{\psi(\pi,v;\theta)}\end{array}\right]\]
where we stack the states $\boldsymbol{v}:=(v_1^{(0)},\dots,v_{K_S}^{(S)})',\boldsymbol{\pi}:=(\pi_1^{(0)},\dots,\pi_{K_S}^{(S)})$ and the mappings
\[\boldsymbol{\phi(\pi,v;\theta)}:=\left(\pi_1^{(0)},\dots,\pi_{K_0}^{(0)},\phi_1^{(1)}\left(\pi_1^{(0)},\dots,\pi_{K_{0}}^{(0)};v_1^{(1)},\dots,v_{K_1}^{(1)}\right),\dots,
\phi_{K_S}^{(S)}\left(\pi_1^{(S-1)},\dots,\pi_{K_{S-1}}^{(S-1)};v_1^{(S)},\dots,v_{K_S}^{(S)}\right)\right)\] and
\[\boldsymbol{\psi(\pi,v;\theta)}:=\left(\psi_1^{(1)}\left(\pi_1^{(1)},\dots,\pi_{K_1}^{(1)};v_1^{(2)},\dots,v_{K_{2}}^{(2)}\right),\dots,
\psi_{K_{S-1}}^{(S-1)}\left(\pi_1^{(S-1)},\dots,\pi_{K_{S-1}}^{(S-1)};v_1^{(S)},\dots,v_{K_S}^{(S)}\right),v_1^{(S)},\dots,v_{K_S}^{(S)}\right)\]

In the following, we will use the more compact notation
\[\mathbf{h}:=\boldsymbol\upsilon(\mathbf{h};\boldsymbol\theta)\]
where $\mathbf{h}:=(\boldsymbol\pi',\boldsymbol{v}')'$ and $\boldsymbol\upsilon:=(\boldsymbol\phi',\boldsymbol\psi')'$. For any given value of $\boldsymbol\theta$ we also denote the solution of this fixed point condition with $\boldsymbol{h}(\boldsymbol\theta)$.

In that notation, we can write the Jacobian of the activation mapping with respect to the hidden states as
\[\boldsymbol{\Upsilon_h}:=\nabla_{\boldsymbol{h}}\boldsymbol\upsilon(\mathbf{h};\boldsymbol\theta)=\left(\frac{\partial}{\partial h_i}
\upsilon_j(\mathbf{h};\boldsymbol\theta)\right)_{ij}\]
By construction $\dim(\boldsymbol\upsilon)=\dim(\boldsymbol{h})$, so that $\boldsymbol{\Upsilon_h}$ is a square matrix. The top diagonal block of this matrix consists of the partial derivatives of $\boldsymbol\phi$ with respect to $\boldsymbol\pi$ which is an upper triangular matrix with all diagonal elements except for the first equal to zero. Similarly, the bottom diagonal block of $\boldsymbol{\Upsilon_h}$ is lower triangular. The bottom left off-diagonal block of $\boldsymbol{\Upsilon_h}$ consisting of partial derivatives of $\boldsymbol\phi$ with respect to $\boldsymbol{v}$ is strictly upper diagonal, and the top right off-diagonal block of that matrix is strictly lower diagonal.

We also denote the Jacobian with respect to activation weights and shape parameters with
\[\boldsymbol{\Upsilon_{\theta}}:=\nabla_{\boldsymbol{\theta}}\boldsymbol\upsilon(\mathbf{h};\boldsymbol\theta)
=\left(\frac{\partial}{\partial \theta_i}\upsilon_j(\mathbf{h};\boldsymbol\theta)\right)_{ij}\]
which is generally not square but sparse since no component of the parameter vector $\boldsymbol\theta$ is shared among multiple elements of the mapping $\upsilon(\mathbf{h};\boldsymbol\theta)$.

Since the objective function $L(\boldsymbol{\theta})$ depends on the parameter only through the hidden states $\boldsymbol{h}$, we can use the chain rule to obtain the gradient
\[\nabla_{\boldsymbol\theta}L(\boldsymbol\theta) = \nabla_{\boldsymbol{\theta}}\mathbf{h}(\boldsymbol\theta)\nabla_{\mathbf{h}}L\]
The gradient $\nabla_{\boldsymbol{h}}L$ is readily available in closed form, with the objective function only depending on states in the bottom and top layer. In contrast, the mapping $\boldsymbol{h}(\boldsymbol\theta)$ is not available in closed form but defined by the fixed point condition
\[\mathbf{h}-\upsilon(\mathbf{h};\boldsymbol\theta) = 0\]
From this fixed-point condition, we can use the implicit function theorem to obtain the derivative
\begin{equation}
\label{implicit_gradient}\nabla_{\boldsymbol{\theta}}\mathbf{h}(\boldsymbol\theta) = \boldsymbol\Upsilon_{\boldsymbol\theta}\left(\mathbf{I}-\boldsymbol\Upsilon_{\mathbf{h}}\right)^{-1}
\end{equation}
assuming the inverse exists, where all derivatives are evaluated at $\boldsymbol{h}(\boldsymbol\theta),\boldsymbol\theta$.

Since the dimension of $\mathbf{h}$ is potentially very large, the main challenge in evaluating this gradient is the computation of the inverse matrix. We propose to sidestep this difficulty by approximating the gradient based on the Neumann representation of the inverse via the infinite series
\[\left(\mathbf{I}-\boldsymbol\Upsilon_{\mathbf{h}}\right)^{-1} = \sum_{q=0}^{\infty}\Upsilon_{\mathbf{h}}^q\]
where the $\mathbf{A}^q$ denotes the $q$-fold product of a square matrix $\mathbf{A}$. If $\mathbf{I}-\boldsymbol\Upsilon_{\mathbf{h}}$ is nonsingular, the series on the right-hand side converges, so that we can approximate its inverse by the finite sum
\begin{equation}
\label{neumann_appx}\left(\mathbf{I}-\boldsymbol\Upsilon_{\mathbf{h}}\right)^{-1} = \sum_{q=0}^{Q}\boldsymbol\Upsilon_{\mathbf{h}}^q + R_Q
\end{equation}
for some $Q<\infty$. If the eigenvalues of $\left(\mathbf{I}-\boldsymbol\Upsilon_{\mathbf{h}}\right)$ are bounded from below by $\underline{\lambda}>0$, then the remainder is bounded by $\|R_Q\|_2\leq\lambda^Q/(1-\lambda)$ under the  spectral matrix norm.

We propose to approximate the inverse according to (\ref{neumann_appx}) where the $q=2,\dots,Q$-fold matrix products are obtained recursively. This approach also exploits that the matrix $\boldsymbol\Upsilon_{\mathbf{h}}$ is sparse for deep networks, where only blocks corresponding to inputs and outputs in adjacent layers can be nonzero.

As a special case, for multilayer feedforward neural networks, the matrix $\Upsilon_{\mathbf{h}}$ is upper triangular, where we can verify that $\boldsymbol\Upsilon_{\mathbf{h}}^{S+1}=0$, so that the approximation in (\ref{neumann_appx}) becomes exact for $Q\geq S$. In particular, for MLP, the backpropagation algorithm by \cite{RHW86} can be interpreted as evaluating the gradient according to (\ref{implicit_gradient}). Specifically, the Neumann series representation to evaluate the matrix inverse using the formula $\left(\mathbf{I}-\boldsymbol\Upsilon_{\mathbf{h}}\right)^{-1} = \sum_{q=0}^{S}\boldsymbol\Upsilon_{\mathbf{h}}^q$ simplifies to the iterative application of the chain rule as under the classical backpropagation algorithm. We can therefore view our approach as a generalization of the backpropagation algorithm to multilayer neural networks \emph{with} feedback, showing that we can control the approximation error from truncating the Neumann series after the first $Q$ summands.

In sum we propose the following algorithm to evaluate the gradient of the objective function:
\begin{enumerate}
\item Obtain and store $\boldsymbol\Upsilon_{\mathbf{h}}$, then
\item compute $J_Q:=\sum_{q=0}^{Q}\boldsymbol\Upsilon_{\mathbf{h}}^q$ recursively.
\item Obtain $\boldsymbol\Upsilon_{\boldsymbol\theta}$ and $\nabla_{\mathbf{h}}L$, and compute the gradient
\[\widehat{\nabla_{\boldsymbol\theta}L} =  \boldsymbol\Upsilon_{\boldsymbol\theta}J_Q\nabla_{\mathbf{h}}L \]
\end{enumerate}
Note that the components $\boldsymbol\Upsilon_{\mathbf{h}}$, $\boldsymbol\Upsilon_{\boldsymbol\theta}$, and $\nabla_{\mathbf{h}}L$ are sparse in the sense that out of the $2\sum_{s=1}^SK_s$ entries along the dimension corresponding to an element of $\boldsymbol{h}$ at most $2\max_sK_s$ can be non-zero. This sparsity can significantly reduce the computational cost of evaluating the matrix products in steps 2 and 3 of this algorithm, especially if the network is deep and narrow.

The resulting vector $\widehat{\nabla_{\boldsymbol\theta}L}$ is only an approximation to the exact gradient $\nabla_{\boldsymbol\theta}L$ unless the matrix product $\boldsymbol\Upsilon_{\mathbf{h}}^{Q+1}$ is equal to zero. This feature is shared with other popular approaches for deep learning models with feedback, for example variational methods for training deep belief networks or deep Boltzmann machines (see e.g. \cite{SHi09}).

Another issue that is shared with other implementations of MLP is the vanishing gradient problem, i.e. that gradients with parameters entering deeper layers in the network will be zero (or near zero) even far from a global optimum. This phenomenon is typically due to near-zero partial derivatives of activation functions with respect to hidden states in subsequent layers. This is especially he ``squashing" activation functions that are natural in the discrete choice context whose derivatives vanish for arguments distant from the origin. This property is not shared by nested CES production functions, which may be less vulnerable to the issue. In the wider literature on deep learning with MLP, the ReLU activation function has become popular for applications in particular since it largely avoids the vanishing gradient problem, however it is more difficult to justify its use for generative models of the type considered in this paper.

\section{Discrete Choics with Flexible Marginals}

\label{sec:GEV_flex_marg_app}

The generalized extreme-value distribution introduced in Section \ref{sec:GEV_model} assumes a generating function that is homogeneous of degree 1 so that the marginal distributions of $\varepsilon_1,\dots,\varepsilon_{M}$ are required to be powers of the extreme-value type I distribution. This model can be made more flexible by allowing for generating functions
\[F(w_1,\dots,w_{M}):=\tilde{F}(\chi_1(w_1),\dots,\chi_{M}(w_{M}))\]
where $\tilde{F}(\cdot)$ is homogeneous of degree one, and the functions $\chi_k:\mathbb{R}_+\rightarrow\mathbb{R}_+$ are continuous and nondecreasing, but otherwise unrestricted. This extended model can generate an arbitrary continuous marginal distribution with c.d.f. $F_k(\varepsilon)$ by choosing
\[\chi_m(w):= -\left[\left.\frac{\partial}{\partial z_m}\tilde{F}(0,\dots,0,z_m,0,\dots,0)\right|_{z_m=\chi_m(w_m)}\right]^{-1}\log G_m(-\log w)\]
We propose to approximate the generating function in two steps, where the linearly homogeneous function $\tilde{F}(z_1,\dots,z_{M})$ is approximated by the cross-nested model (\ref{CNL_recursion}), and the component-wise functions $\chi_m(w)$ by a polynomial
\[\hat{\chi}_{mK}(y):=\sum_{l=1}^K \beta_{ml} w^l\]
By Weierstrass' theorem, for an appropriate choice of $K$ and the coefficients $\beta_{ml}$, $\hat{\chi}_{mK}(w)$ approximates $\chi_m(w)$ uniformly on an arbitrarily chosen compact subset of $\mathbb{R}_+$. This approximation consists of two nested layers
\[ \hat{\chi}_{mK}(w):=F_m^{(2)}\left(F_{m1}^{(1)}(w),\dots,F_{mK}^{(1)}(w)\right)\]
where $F_m^{(2)}\left(w_1,\dots,w_K\right):=\sum_{l=1}^K\beta_{ml}^{(2)}w_l$ and $F_{ml}^{(1)}(z):=z^l$ are both special cases of
\[F_m^{(s)}(w_1,\dots,w_{K_s}):=\left(\sum_{l=1}^{K_{s+1}}\left(\beta_{ml}^{(s)}w_l^{\tau_m^{(s)}}\right)^{\varrho_m^{(s)}}\right)^{\frac{1}{\varrho_m^{(s)}}}\]
where $\tau_m^{(s)}$ may be different from one.

Note that this activation function is homogeneous of degree $\tau_m^{(s)}$, so that we first need to generalize results for the nesting functions to nonlinearly homogenous functions. We can first establish the following generalization of Theorem 1 in \cite{McF78}:

\begin{lem}\label{CNL_nonlin_homog_lem}
Suppose that the assumptions of Theorem 1 in \cite{McF78} hold, but that the generating function $F(w_1,\dots,w_{M})$ is of arbitrary positive degree of homogeneity, $F(\lambda w_1,\dots,\lambda w_{M})=\lambda^{\tau}F(w_1,\dots,w_{M})$. Then
\[P\left(\max_{m}U_m=U_k\right) = \frac{e^{U_k^*}\frac{\partial}{\partial w_k}F\left(e^{U_1^*},\dots,e^{U_{M}^*}\right)}{\tau F\left(e^{U_1^*},\dots,e^{U_{M}^*}\right)}\]
\end{lem}

See the appendix for a proof. Moreover, the conditional choice probabilities resulting from the nested model with stage-wise generating functions
\[F_k^{(s)}(w_1,\dots,w_{K_{s+1}})
=\left(\sum_{l=1}^{K_{s+1}}\left(\beta_{kl}^{(s)}w_l^{\tau_k^{(s)}}\right)^{\varrho_k^{(s)}}\right)^{\frac{1}{\varrho_k^{(s)}}}\]
can be written in terms of inclusive values determined by the recursion
\begin{eqnarray}
\nonumber v_k^{(S)}&:=&e^{U_k^*}\\
\nonumber
v_k^{(s)}&:=&\psi_k^{(s)}(v_1^{(s+1)},\dots,v_{K_{s+1}}^{(s+1)})
\end{eqnarray}
where
\begin{equation}
\psi_k^{(s)}(v_1^{(s+1)},\dots,v_{K_{s+1}}^{(s+1)}):=
\log\left(\left[\sum_{l=1}^{K_{s+1}}\left(\beta_{kl}^{(s)}e^{\tau_k^{(s)}v_l^{(s+1)}}\right)^{\varrho_k^{(s)}}\right]^{1/\varrho_k^{(s)}}\right)
\end{equation}
Applying this recursion to the mapping $\hat{\chi}_k(\cdot)$ we obtain an approximation
\[\hat{F}\left(e^{U_1^*},\dots,e^{U_{M}^*}\right) = \tilde{F}\left(\hat{\chi}_1(e^{U_1^*}),\dots,\hat{\chi}_M(e^{U_{M}^*})\right)\]

It is important to notice that the approximating function $\hat{\chi}_m(w)$ is generally not nonnegative or nondecreasing at each value of $w$, so the iteration (\ref{nest_inclusive_val_psi}) would generally require that arguments be trimmed at some small positive number. However since the approximand is positive and nondecreasing, this would not affect the rate of approximation under Proposition \ref{CNL_nested_universal_prp}. Furthermore, the bottom layer of this network generates nests consisting of a single alternative, which will therefore be ``selected" with probability one conditional on reaching that nest. Nevertheless, these intermediate nests are not irrelevant since they alter the inclusive values that get passed on to the next layer. Again following the representation for nested Logit in \cite{BLe85} and \cite{Gal21}, this approach can be interpreted as an approximation of the marginal distributions of taste shocks by a weighted sum of independent draws from an extreme value distribution.

\section{Proofs}

\subsection{Proof of Proposition \ref{CES_prod_universal_prp}}
We fix the second-stage technology for each output at
\[F_m^{(2)}(w_1^{(1)},\dots,w_{K_1}^{(1)}):= \sum_{l=0}^{K_1}\beta_{ml}^{(2)}w_l^{(2)}\]
where $\beta_{ml}^{(2)}=1$ if $(l-m)/M$ is an integer, and zero otherwise. This production function satisfies (\ref{CES_activation_fct}) with parameter values $\tau^{(2)}=\varrho^{(2)}=1$ and $\beta_{ml}^{(2)}=1$ for all $l=1,\dots,K_1$ and $m=1,\dots,M$.

Next we fix a point $(\mathbf{y}_q',\mathbf{w}_q')'$ on the boundary of $\mathcal{Y}_0$ and choose $M$ first-stage technologies
\[\tilde{F}_{qm}^{(1)}(w_1^{(0)},\dots w_{K_0}^{(0)}) = \min\left\{\tilde{\beta}_{qm0},\tilde{\beta}_{qm1}w_1^{(0)},\dots,\tilde{\beta}_{qmK_0}w_K^{(0)}\right\}\]
for $m=1,\dots,M$, where we choose $\tilde{\beta}_{qml}^{(1)}:=\frac{y_{qm}}{w_{ql}/M}$ where $y_{qm}$ denotes the $m$th component of $\mathbf{y}_q$, and $w_{ql}$ denotes the $l$th component of $\mathbf{w}_q$. These technologies transforms equal fractions $\frac1M$ of the input into $y_{q1},\dots,y_{qM}$ units of the intermediate output.

We can apply this construction to a set of support points $(\mathbf{y}_q',\mathbf{w}_q')'$ with $q=1,\dots,Q$. To simplify notation we change the order of enumeration for the first stage technologies and let
\[F_{M(q-1) + m}^{(1)}(w_1^{(0)},\dots,w_{K_0}^{(0)}):=\tilde{F}_{qm}^{(1)}(w_1^{(0)},\dots,w_{K_0}^{(0)})\]
where we set $\beta_{(M(q-1)+m)l}^{(1)}:=\tilde{\beta}_{qml}^{(1)}$ and the number of intermediate goods in the first stage equals $K_1 = MQ$.



By construction the technology set $\mathcal{Y}_{K_1}$ resulting from this two stage technology includes the points $(\mathbf{y}_q',\mathbf{w}_q')'$ for each $q=1,\dots,Q$. By inspection, we also have $0\in\mathcal{Y}_{K_1}$. The intermediate technologies also satisfy free disposal by assumption, so that $(\mathbf{y}_q,\mathbf{w}_q+\mathbf{t})$ and $(\mathbf{0},\mathbf{t})$ are also in $\mathcal{Y}_{K_1}$ for any $\mathbf{t}\geq0$.

Furthermore, we argue that any convex combination of these points is included in $\mathcal{Y}_{K_1}$ as well: for the point $\sum_{q=1}^Q\lambda_q(\mathbf{y}_q',\mathbf{w}_q')'$ with $\sum_{q=1}\lambda_q\leq 1$, we consider a production plan that employs a quantity $\lambda_q\mathbf{w}_q/M$ for production with the intermediate technology $\phi_{M(q-1)+m}(\cdot)$, and the resulting output $\mathbf{w}_{M(q-1)+m}^{(1)}$ for production of the output $y_m$. Since each first-stage technology $\phi_{M(q-1) + m}^{(1)}(\cdot)$ has constant returns to scale for inputs $\mathbf{w}^{(0)}\leq \mathbf{w}_q/M$, the resulting output is $\sum_{q=1}^Q\lambda_q\mathbf{y}_q$ so that the convex combination $\sum_{q=1}^Q\lambda_q(\mathbf{y}_q',\mathbf{x}_q')'\in\mathcal{Y}_{K_1}$.

Next, consider a point $(\mathbf{\tilde{y}}',\mathbf{\tilde{w}}')'$ such that for any $\lambda_1,\dots,\lambda_Q\geq0$ with $\sum_q\lambda_q\leq1$, we have $\sum_q\lambda_q\mathbf{\tilde{y}}_q<\mathbf{\tilde{y}}$ or $\sum_q\lambda_q\mathbf{\tilde{x}}_q>\mathbf{\tilde{x}}$. Then for any production plan there must be $m\leq M$ and $q\leq Q$ such that
\[F_{M(q-1)+m}^{(1)}(w_{1q}^{(0)},\dots,w_{K_0q}^{(0)}) = \min_l\left\{\frac{y_{qm}}{w_{ql}/M}w_{lq}^{(0)}\right\}<\tilde{y}_m,\]
so that the point $(\mathbf{\tilde{y}}',\mathbf{\tilde{w}}')'$ is infeasible. Hence $\mathcal{Y}_{K_1}$ excludes all points outside the convex hull of $(\mathbf{y}_q',\mathbf{w}_q'+\mathbf{t})'$, for $q=1,\dots,Q$ and $\mathbf{t}\in\{0,\infty\}^{K}$. 

Hence the technology set $\mathcal{Y}_{K_1}$ forms a polytope with a vertex set consisting of the origin and points $(\mathbf{y}_q',\mathbf{w}_q'+\mathbf{t})'$, for $q=1,\dots,Q$ and $\mathbf{t}\in\{0,\infty\}^{K}$. Since $\mathcal{Y}_0$ is convex as well, and the choice of points $(\mathbf{y}_q',\mathbf{x}_q')$, $q=1,\dots,Q$ was arbitrary, the approximation can be made arbitrarily close by choosing $Q$ large enough. The rate of approximation under the Hausdorff metric, $d_H(\mathcal{Y}_0,\mathcal{Y}_{\mathbf{K}}^S)$ given $K_1$ follows directly from the main theorem for the approximation of a convex body  using convex polyhedra in \cite{BIv76}. Since $\mathcal{C}$ is rectangular, we can assume without loss of generality that the vertices of $\mathcal{Y}_{K_1}$ contain the input/output combinations corresponding to the $2^K$ vertices of $\mathcal{C}$, so that the projection of $\mathcal{Y}_{K_1}$ onto its first $K$ components contains all of $\mathcal{C}$.

Now let the set $\mathcal{Y}_0(\mathbf{x})$ denote the intersection of a $\mathcal{Y}_0$ with the subspace $\left\{(\mathbf{x},\mathbf{u}):\mathbf{u}\in\mathbb{R}^M\right\}$. Since $\mathcal{Y}_0(\mathbf{x})\cap\mathbb{R}_+^M\neq\emptyset$, $\mu_0(\mathbf{u},0)\geq0$ for all $\mathbf{u}\in\mathbb{R}_+^M$. Furthermore, free disposal implies that $\mathcal{Y}_0(\mathbf{x})\subset\mathcal{Y}_0(\mathbf{x}')$ whenever $\mathbf{x}\leq\mathbf{x}'$, so that $\mu_0(\mathbf{u},\mathbf{x})$ is nondecreasing in $\mathbf{x}$ with respect to the component-wise order. In particular for compact $\mathcal{C}\subset\textnormal{int}\mathbb{R}_+^{K}$, the partial derivatives of $\mu_0(\mathbf{u},\mathbf{x})$ are bounded over $\mathbf{x}\in\mathcal{C}$, implying that $\mu_0(\mathbf{u},\mathbf{x})$ is Lipschitz on $\mathcal{C}$. Therefore $d_H(\mathcal{Y}_0,\mathcal{Y}_{\mathbf{K}}^S)<\delta$ implies $\sup_{\mathbf{x}\in\mathcal{C},\mathbf{u}\in\mathbb{R}^M}\left|\mu_0(\mathbf{u},\mathbf{x})-\mu_{\mathbf{K}}^S(\mathbf{u},\mathbf{x})\right|<L_1\delta$ for some $L_1<\infty$. Since $\delta>0$ was arbitrary, this establishes the claim\qed



\subsection{Proof of Proposition \ref{CNL_nested_universal_prp}} Fix $\delta>0$. In light of \cite{McF78}'s characterization of conditional choice probabilities (\ref{mcf_gev_ccp}), we will give a constructive proof that a two-layer network with $K_1$ nodes in the top layer suffices to approximate the level and first partial derivatives of $F_0(w_1,\dots,w_m)$ up to an error not exceeding $\delta>0$.

We first show that the network can approximate the subgraph of $F(\cdot)$ on $\mathcal{C}$: note that Assumption \ref{GEV_generating_ass} implies that $F_0(\mathbf{w})$ is convex on $\mathbb{R}_+^{M}$: Since all second-order cross-partial derivatives $\frac{\partial^2}{\partial w_j \partial w_k}F_0(w_1,\dots,w_{M})$ are negative for any $j\neq k$, the function is submodular with respect to the component-wise order on $\mathbb{R}_+^{M}$ (see \cite{Top78}). Since $F(\cdot)$ is furthermore homogeneous of degree 1 on $\mathbb{R}_+^M$, it then follows from Theorem 3 in \cite{MMo08} that the function is also quasiconvex (a previous version of that result was previously given as Theorem 54.1 in \cite{Cho54}).

Furthermore, any homogeneous function is homothetic, so that it is sufficient to show that we can approximate the lower contour set of $F$ for one particular non-zero value - without loss of generality, we will focus on one lower contour set for $F_0$,
\[LC(\lambda):=\left\{\mathbf{w}\in\mathbb{R}^{M}:F_0(\mathbf{w})\leq\lambda\right\}\]
for some arbitrarily chosen level $\lambda>0$, where we choose $\lambda=1$. Since $F_0$ is quasi-convex, the set $LC(\lambda)$ is convex.

We therefore first show that we can construct a neural network resulting in a linearly homogeneous function $F_{\mathbf{K}}^S(\mathbf{w})$ such that the lower contour set $LC_{\mathbf{k}}^S(\lambda):=\left\{\mathbf{w}\in\mathbb{R}^{M}:F_{\mathbf{K}}^S(\mathbf{w})\leq\lambda\right\}$ approximates $LC(\lambda)$ according to
\[d_H(LC_{\mathbf{K}}^S(\lambda),LC(\lambda))\leq\delta\]
where $d_H(\mathcal{A},\mathcal{B})$ denotes the Hausdorff distance between sets $\mathcal{A}$ and $\mathcal{B}$.

We can verify that the class (\ref{CNL_generating_fct}) include the functions $\beta_1w_1+\dots+\beta_{M}w_{M}$ and $\max\left\{\beta_1w_1,\dots,\beta_{M}w_{M}\right\}$, corresponding to $\varrho=1$ and $\varrho=+\infty$, respectively. This allows us to define a four-layer model according to (\ref{CNL_recursion}), where the top layer $s=3$ consists of the $K_3=M$ neurons with activations $w_m^{(3)}=\exp\left\{U_m^*\right\}$. Layer $s=2$ consists of $K_2$ nodes with generating functions \[F_k^{(2)}:=\max\left\{\beta_{k1}^{(2)}w_1,\dots,\beta_{kM}^{(2)}w_{M}\right\}.\]
Layer $s=1$ consists of $K_1=\binom{K_2}{M}$ neurons, indexed by all subsets $P_k$ of $M$ out of $K_1$ neurons, where the $k$th neuron is identified with the generating function
\[F_k^{(1)}:=\sum_{l=1}^{K_2}\beta_{kl}^{(1)} w_l^{(2)}\]
where $\beta_{kl}^{(1)}:=\frac1M\dum\{l\in P_k\}$. The bottom (root) layer $s=0$ consists of a single node, $K_0=1$,
\[F_1^{(0)}:=\max\left\{w_1^{(1)},\dots,w_{K_1}^{(1)}\right\}\]
Note that in this construction, only the weights in layer 2 are varying freely.

The composition of these mappings defines the boundary of a convex polytope with $K_2$ vertices and $O(K_2)$ faces in $\mathbb{R}^{M+1}$. In particular, the maximum in $F_{1}^{(0)}(\cdot)$ is attained only at a subset of $O(K_2)$ nodes, layer 1 could be replaced by a layer containing only $O(K_2)$ nodes that would be fully determined by the weights in layer 2. Those $O(M)$ nodes correspond to a tesselation such that no other node is contained in the convex hull of $P_k$.

The locations of the vertices are determined by the coefficients in layer 2, which we choose as follows: let $B:=\max_{\mathbf{w}\in\mathcal{C}}\max\{w_1,\dots,w_M\}$ and $L:=\lceil\frac{B}{\delta}\rceil$, the smallest integer greater than $\frac{B}{\delta}$. We then construct a grid of $K_2:=L^{M-1}$ points $(w_{l_1,1},\dots,w_{l_{M-1},M-1})=(l_1,\dots,l_{M-1})\delta$ with $l_1,\dots,l_{M-1}\in\{1,\dots,L\}$, and we also let $w_{l_1,\dots,l_{M-1},M},:=\min\left\{z:F(w_{l_1,1},\dots,w_{l_{M-1},M-1},z)\geq \lambda\right\}$. We then define
\[\mathbf{w}_{k(l_1,\dots,l_{M-1})}:=(w_{l_1,1},\dots,w_{l_{M-1},M-1},w_{l_1,\dots,l_{M-1},M})\]
where we index grid points by $k(l_1,\dots,l_{M-1}) = l_1 + l_2L + \dots + l_{M-1}L^{M-2}$. We will then use this grid as support points on the boundary of the lower contour set of $F_0(w_1,\dots,w_{M})$.

Specifically, we choose the parameters in (\ref{CNL_generating_fct}) for neurons in the first layer so that $F_k^{(2)}(w_{k,1},\dots,w_{k,M})=1$ by setting $\varrho_k^{(2)}=\infty$ and
\[\beta_{kj}^{(2)}:=\frac1{w_{k,j}}\]
for each $k=1,\dots,K_1$. Since $F(w_1,\dots,w_{M})$ is non-decreasing and linearly homogeneous, $F_k^{(2)}(w_1,\dots,w_{M})\leq F(w_1,\dots,w_{M})$ for all $w_1,\dots,w_{M}$, so that the lower contour set
\[LC_k^{(2)}(\lambda):=\{\mathbf{w}\in\mathbb{R}^{M}:F_k^{(2)}(\mathbf{w})\leq\lambda\}\subset LC(1).\]
noting that $LC(1)$ is convex. Similarly, $LC_k^{(1)}(\lambda)$ is the convex hull of $\bigcup_{m\in P_k} LC_{l_m}^{(2)}(\lambda)$ which is also contained in $LC(\lambda)$ by convexity. Finally, for $F_1^{(0)}:=\max\{w_1,\dots,w_{K_1}\}$, we have that
$LC_{1}^0(1)$ is the union of $LC_1^{(1)}(\lambda),\dots,LC_{K_1}^{(1)}(\lambda)\subset LC(\lambda)$. In particular, $LC_1^{(0)}(1)\subset LC(1)$, so that it remains to be shown that $d(\mathbf{w},LC_{\mathbf{k}}^S(1))\leq\delta$ for any point in $LC(1)$.

To this end, fix an arbitrary point $\tilde{\mathbf{w}}=(\tilde{w}_1,\dots,\tilde{w}_{M})\in LC$, where the adjacent grid points for the first $M-1$ coordinates are determined according to $w_{l_j,j}\leq \tilde{w}_j\leq w_{l_j+1,j}$. By construction, each of the points $\mathbf{w}_k$ is on the boundary of $LC(1)$, so that by monotonicity of $F(\mathbf{w})$, $w_{l_1,\dots,l_{M-1},M}\leq \tilde{w}_{M}$. Moreover, monotonicity of $F_{k(l_1,\dots,l_{M-1})}^{(2)}(w_1,\dots,w_{M})$ in each argument also implies that $(w_{l_1,1},\tilde{w}_2,\dots,\tilde{w}_{M})\in LC_{\mathbf{k}}^S(1)$, so that
\[d(\tilde{\mathbf{w}},LC_{\mathbf{k}}^S(1))\leq |\tilde{w}_1-w_{l_1,1}|\leq\delta.\]
so that indeed $d_H(LC(\lambda),LC_{\mathbf{k}}^S(\lambda))\leq\delta$ as claimed.

Since $\mathcal{C}$ is compact and $F_0(\mathbf{w})$ is continuous, $\lambda^*:=\max_{\mathbf{w}\in\mathcal{C}}F(\mathbf{w})<\infty$ is well-defined and finite. Furthermore, by linear homogeneity and monotonicity, $LC(\lambda) = \lambda LC(1)$ and $LC_{\mathbf{k}}^S(\lambda)=\lambda LC_{\mathbf{k}}^S(1)$, so that for every $\lambda\in[0,\lambda^*]$,
\[d_H(LC(\lambda),LC_{\mathbf{k}}^S(\lambda))=\frac{\lambda}{\lambda^*}d_H(LC(\lambda^*),LC_{\mathbf{k}}^S(\lambda^*))\leq\delta\]
Furthermore, by construction \[F_{\mathbf{K}}^3(\mathbf{w}):=F_1^{(0)}\left(F_1^{(1)}(F_1^{(2)}(\mathbf{w}),\dots,F_{K_2}^{(2)}(\mathbf{w})),\dots,
F_{K_1}^{(1)}(F_1^{(2)}(\mathbf{w}),\dots,F_{K_2}^{(2)}(\mathbf{w}))\right)\leq F_0(w_1,\dots,w_{M}),\] so that the maximal Hausdorff distance between the lower contour sets corresponding to levels $\lambda\in[0,\lambda^*]$ is an upper bound for the Hausdorff distance between the subgraphs of $F_{\mathbf{K}}^3(w_1,\dots,w_M)$ and $F(w_1,\dots,w_{M})$, respectively, intersected with $\mathcal{C}\times\mathbb{R}$. The rate of approximation with regards to the generating function $F_0(w_1,\dots,w_M)$ then follows directly from applying the main theorem for the approximation of a convex body using convex polyhedra in \cite{BIv76} to the lower contour sets of $F_0(\mathbf{w})$.

Next we need to apply a similar argument to bound the approximation error to the first partial derivatives, $\frac{\partial}{\partial w_m}F_0(w_1,\dots,w_M)$, to obtain rates for conditional choice probabilities according to (\ref{mcf_gev_ccp}). Note that the second part of Assumption \ref{GEV_generating_ass} implies that the first partial derivative is non-negative and a convex function of $w_1,\dots,w_M$. Furthermore, linear homogeneity of the generating function implies homogeneity of degree zero of its first partial derivatives, so that without loss of generality it suffices to approximate $frac{\partial}{\partial w_m}F(\tilde{w}_1,\dots,\tilde{w}_M)$ for values of $\mathbf{w}$ in the simplex $\mathcal{S}:=\left\{\mathbf{w}\in\mathbb{R}_+^M:\sum_{k=1}^Mw_k = 1\right\}$. Furthermore, since by assumption the test set $\mathcal{C}$ is a compact subset of the interior of the positive orthant and $\frac{\partial}{\partial w_m}F_0(\mathbf{w})\geq0$ and $\frac{\partial^2}{\partial w_m^2}F_0(\mathbf{w})\leq0$, it follows that $\left|\frac{\partial}{\partial w_m}F_0(\mathbf{w})\right|\leq B_1$ for some $B_1<\infty$.

For the approximating network, note that $F_{\mathbf{K}}^S(\cdot)$ is piecewise linear on each of the partition elements of $\mathbb{R}^M$ corresponding to the projection of the faces of the polytope approximating $LC(\lambda)$ onto their first $M$ coordinates. In particular, the partial derivative $\frac{\partial}{\partial w_m}F_{\mathbf{K}}^S(\cdot)$ is constant within each set of that partition. By the mean value theorem, we furthermore have $\frac{\partial}{\partial w_m}F_{\mathbf{K}}^S(\tilde{w}_1,\dots,\tilde{w}_M) = \frac{\partial}{\partial w_m}F_0(\tilde{w}_1,\dots,\tilde{w}_M)$ for some value of $(\tilde{w}_1,\dots,\tilde{w}_M)$ in that set. Finally, by inspection $F_{\mathbf{K}}^S(\cdot)$ is also homogeneous of degree 1, so that its directional  partial derivatives are homogenous of degree zero, as for the approximand $F_0(\mathbf{w})$.

For an arbitrary choice of $\delta_0>0$ we can now partition the range of $\frac{\partial}{\partial w_m}F_0(w_1,\dots,w_M)$ into $Q = \lfloor B_1\delta_0^{-1}\rfloor$ intervals of the form $[\lambda_q-\delta_0,\lambda_q]$ with partition points $\lambda_1=\delta_0,\dots,\lambda_Q=B_1$. Since $\frac{\partial}{\partial w_m}F(w_1,\dots,w_M)$ is a convex function, any lower contour set $LC_{m}(\lambda_q):=\left\{\mathbf{w}\in\mathbb{R}^M:\frac{\partial}{\partial w_m}F(w_1,\dots,w_M)=\lambda_q\right\}$ is a convex set in $\mathbb{R}^M$ for each $\lambda_q$. Using again the main theorem in \cite{BIv76} the intersection of $LC_{m}(\lambda_q)$ with the $(M-1)$-dimensional simplex $\mathcal{S}$ can be approximated with the intersection of a polytope with $K_{2m}=c_m(R,M)\delta_0^{-{M-2}2}$ vertices with $\mathcal{S}$. We can therefore approximate all $Q$ lower contour sets jointly over a tesselation with $QK_{2m}(\lambda_q) = c_m(R,M)\delta_0^{-{M}2}$ vertices.

Now, for a given value $\mathbf{w}$, $\frac{\partial}{\partial w_m}F_{\mathbf{K}}^S(\mathbf{w})$ is equal to $\frac{\partial}{\partial w_m}F_0(\tilde{\mathbf{w}})$ for a point $\tilde{\mathbf{w}}$ that is part of the same partition set. Now choose $\lambda,\tilde{\lambda}\in\{\lambda_1,\dots,\lambda_Q\}$ such that $\tilde{\lambda}-\delta_0\leq\frac{\partial}{\partial w_m}F_0(\tilde{\mathbf{w}})\leq\tilde{\lambda}$ and $\lambda-\delta_0\leq\frac{\partial}{\partial w_m}F_0(\mathbf{w})\leq\lambda$, where we assume w.l.o.g. that $\lambda\leq\tilde{\lambda}$. If $\lambda=\tilde{\lambda}$, then by choice of $\lambda,\tilde{\lambda}$, we have $\left|\frac{\partial}{\partial w_m}F_0(\tilde{\mathbf{w}})-\frac{\partial}{\partial w_m}F_0(\mathbf{w})\right|\leq 2\delta_0$.

For the case $\lambda<\tilde{\lambda}$, we have by construction that either $d(\tilde{w},LC_{m}(\lambda))\leq\delta_0$ or $d(w,UC_m(\tilde{\tilde{\lambda}-\delta_0}))\leq\delta_0$ for the upper contour set $UC_m(\tilde{\tilde{\lambda}-\delta_0}):=\left\{\mathbf{w}\in\mathbb{R}^M:\frac{\partial}{\partial w_m}F_0(\mathbf{w})\geq \tilde{\lambda}
-\delta_0\right\}$. Since $\mathcal{C}$ is compact and third partial derivatives are monotone by Assumption \ref{GEV_generating_ass}, there exists $B_2<\infty$ such that $\left|\frac{\partial^2}{\partial w_m\partial w_l}F_0(\mathbf{w})\right|\leq B_2$ for every $l=1,\dots,M$ and $\mathbf{w}\in\mathcal{C}\cap\mathcal{S}$. It therefore follows that $\left|\frac{\partial}{\partial w_m}F_0(\tilde{\mathbf{w}})-\frac{\partial}{\partial w_m}F_0(\mathbf{w})\right|\leq (B_2+2)\delta_0$. Hence choosing $\delta_0 = \delta/(B_2+2)$ we can approximate $\frac{\partial}{\partial w_m}F_0(\mathbf{w})$ as a function of $\mathbf{w}$ uniformly over $\mathcal{C}$ up to an error no greater than $\delta$.

Taking the union of the $K_2:=c(R,M)\delta^{-{M-1}2}+\sum_{m=1}^Mc_m(R,M)\left(\frac{\delta}{B_2+2}\right)^{-{M}2}=:\bar{c}(R,M)\delta^{-\frac{M}2}$ support points for approximating the level of $F(\mathbf{w})$ and the partial derivatives $\frac{\partial}{\partial w_k}F(\mathbf{w})$, we obtain a network that is capable of approximating the level and all partial derivatives jointly up to an error no greater than $\delta$. Since $F_0(\mathbf{w})$ is bounded away from zero on $\mathcal{C}$, the mapping of levels and partial derivatives to
$\frac{w_m\frac{\partial}{\partial w_m}F_0(\mathbf{w})}{F_0(\mathbf{w})}$ is Lipschitz continuous with Lipschitz constant $L_1<\infty$, say, so that $\|\mu_{\mathbf{K}}^S(y,x)-\mu_0(y,x)\|\leq L_1\delta$, establishing the claim\qed

\subsection{Proof of Lemma \ref{ccp_phi_lem}}

From Theorem 1 in \cite{McF78},
\[P\left(\max_j U_{j} = U_{k}\right) = \frac{e^{U_{k}^*}\frac{\partial}{\partial w_k}F\left(e^{U_{1}^*},\dots,e^{U_{M}^*}\right)}
{F\left(e^{U_{1}^*},\dots,e^{U_{M}^*}\right)}=:\pi^{(S)}_k\]
where the generating function $F(\cdot)$ was defined by the recursion (\ref{CNL_recursion}). The derivative $\frac{\partial}{\partial w_K}F(\cdot)$ can then be obtained via the chain rule
\begin{equation}\label{app_gev_chain_rule}\frac{\partial}{\partial w_k}F(\cdot) = \sum_{k_{S-1}=1}^{K_{S-1}}\dots \sum_{k_1=1}^{K_1} \frac{\partial}{\partial w_{k_{S-1}}}F_{1}^{(S)}
\cdot\dots\cdot \frac{\partial}{\partial w_{k}}F_{k_1}^{(1)}\end{equation}
and the partial derivatives of (\ref{CNL_generating_fct}) are
\begin{eqnarray}
\nonumber \frac{\partial}{\partial w_{l}}F_{k}^{(s)}\left(e^{v_1^{(s-1)}},\dots,e^{v_{K_{s-1}}^{(s-1)}}\right)
&=&\left[\sum_{m=1}^{K_{s-1}}\left(\beta_{km}^{(s)}e^{v_m^{(s-1)}}\right)^{\varrho_k^{(s)}}\right]^{\frac{1}{\varrho_k^{(s)}}-1}
\left(\beta_{kl}^{(s)}e^{v_l^{(s-1)}}\right)^{\varrho_k^{(s)}}e^{-v_l^{(s-1)}}\\
\nonumber&=&e^{v_k^{(s)}}\frac{\left(\beta_{kl}^{(s)}e^{v_l^{(s-1)}}\right)^{\varrho_k^{(s)}}}
{\sum_{m=1}^{K_{s-1}}\left(\beta_{km}^{(s)}e^{v_m^{(s-1)}}\right)^{\varrho_k^{(s)}}}e^{-v_l^{(s-1)}}
\end{eqnarray}
where the second equality follows from (\ref{nest_inclusive_val_psi}). Furthermore, (\ref{nest_inclusive_val_psi}) also implies that
\[F\left(e^{U_{1}^*},\dots,e^{U_{M}^*}\right)= e^{v_1^{(S)}}\]
Hence, we can simplify the products
\[\frac{\partial}{\partial w_{k_{S-1}}}F_{1}^{(S)}
\cdot\dots\cdot \frac{\partial}{\partial w_{j}}F_{k_1}^{(1)}
=\frac{\left(\beta_{k_Sk_{S-1}}^{(S)}e^{v_{k_{S-1}}^{(S-1)}}\right)^{\varrho_{k_S}^{(S)}}}
{\sum_{m=1}^{K_{S-1}}\left(\beta_{k_Sk_{S-1}}^{(S)}e^{v_m^{(S-1)}}\right)^{\varrho_{k_S}^{(S)}}}\cdot\dots
\frac{\left(\beta_{k_1j}^{(1)}e^{U_j^*}\right)^{\varrho_{k_1}^{(1)}}}
{\sum_{q=1}^{M}\left(\beta_{k_1q}^{(1)}e^{U_q^*}\right)^{\varrho_{k_1}^{(1)}}}
F\left(e^{U_{1}^*},\dots,e^{U_{M}^*}\right)e^{-U_j^*}\]
Substituting this into (\ref{app_gev_chain_rule}), we obtain
\[\pi_k^{(S)} =  \sum_{k_{S-1}=1}^{K_{S-1}}\dots \sum_{k_1=1}^{K_1}
\prod_{s=1}^S\frac{\left(\beta_{k_sk_{s-1}}^{(s)}e^{v_{k_{s-1}}^{(s-1)}}\right)^{\varrho_{k_s}^{(s)}}}
{\sum_{m=1}^{K_{s-1}}\left(\beta_{k_sk_{s-1}}^{(s)}e^{v_m^{(s-1)}}\right)^{\varrho_{k_s}^{(s)}}}\]
Note that the same expression can be obtained alternatively via the recursion (\ref{nest_pi_recursion}) with the mapping (\ref{nest_ccp_phi}) given the inclusive values, establishing the claim\qed

\subsection{Proof of Lemma \ref{CNL_nonlin_homog_lem}} Notice that most parts of the original proof do not require homogeneity of degree 1 and therefore continue to go through. Rather we only need to update the calculation for $P_i$ on p.11. Noting that the partial derivative of a function that is homogeneous of degree $\tau$ is homogeneous of degree $\tau-1$, we have for $k=1$
\begin{eqnarray}
\nonumber P\left(\max_{j}U_j=U_1\right)&=&\int_{-\infty}^{\infty}\frac{\partial}{\partial \varepsilon_1}G(\varepsilon,U_1^*-U_2^*+\varepsilon,\dots,U_1^*-U_{M}^*+\varepsilon)d\varepsilon\\
\nonumber&=&\int_{-\infty}^{\infty}e^{-\varepsilon}\frac{\partial}{\partial w_1} F\left(e^{-\varepsilon},e^{-\varepsilon-U_1^*+U_2^*},\dots,e^{-\varepsilon-U_1^*+U_{M}^*}\right)\\
\nonumber&&\exp\left\{-F\left(e^{-\varepsilon},e^{-\varepsilon-U_1^*+U_2^*},\dots,e^{-\varepsilon-U_1^*+U_{M}^*}\right)\right\}d\varepsilon\\
\nonumber&=&\int_{-\infty}^{\infty}e^{-\tau\varepsilon-(\tau-1)U_1^*}\frac{\partial}{\partial w_1}F\left(e^{U_1^*},\dots,e^{U_{M}^*}\right)
\exp\left\{-e^{-\tau(\varepsilon+U_1^*)}F\left(e^{U_1^*},\dots,e^{U_{M}^*}\right)\right\}d\varepsilon\\
\nonumber&=& \frac{e^{U_1^*}\frac{\partial}{\partial w_1}F\left(e^{U_1^*},\dots,e^{U_{M}^*}\right)}{\tau F\left(e^{U_1^*},\dots,e^{U_{M}^*}\right)}
\end{eqnarray}
evaluating the integral using the change of variables $t = e^{-\tau(\varepsilon +U_1^*)}$. The calculation for $k\neq1$ is completely analogous so that the conclusion follows as claimed\qed

\subsection*{Proof of Proposition \ref{sparse_appx_prp}} Fix $\delta>0$. By assumption the functions $F_k^{(s)}(\mathbf{w}^{(s)})$ satisfy the conditions for the target function in Proposition \ref{CNL_nested_universal_prp}, and the approximating two-layer perceptron as a function of the $d^*$ non-trivial inputs can be replicated by the top two layers of the network (SP-CES). Hence it follows from Proposition \ref{CNL_nested_universal_prp} that each $F_k^{(s)}(\mathbf{w}^{(s)})$ can be approximated at an error not exceeding $\delta$ by including $Q_k^{(s)}= c_1(R,d^*)\delta^{-\frac{d^*}{2}}$ neurons in the middle layer. Since the stage functions are uniformly Lipschitz and $S$ is finite, it follows that $Q_k^{(s)}$ can be chosen to be equal to a common value $Q$ of the same order in $\delta$, but with a constant $c(R,S,d^*)<\infty$ depending on $c_1(R,d^*)$, $S$, and the common Lipschitz constant\qed

\subsection*{Proof of Proposition \ref{separable_appx_prp}} Fix $\delta>0$. By assumption the functions $g_{kl}^{(s)}(w_l^{(s-1)})$ and $f_k^{(s)}(z)$ satisfy the conditions for the target function in Proposition \ref{CES_prod_universal_prp} with $K=1$, and the respective approximating two-layer perceptrons can be replicated by the second and fourth layers of the network (4L-CES). Hence it follows from Proposition \ref{CES_prod_universal_prp} that each $F_k^{(s)}(\mathbf{w}^{(s)})$ can be approximated at an error not exceeding $\delta$ by including $Q_k^{(s)}= c_1(R)\delta^{-\frac{1}{2}}$ neurons in the middle layer. Since the stage functions are uniformly Lipschitz and $S$ and $d$ are finite, it follows that $Q_k^{(s)}$ can be chosen to be equal to a common value $Q$ of the same order in $\delta$, but with a constant $c(R,S,K)<\infty$ depending on $c_1(R)$, $S$, and the common Lipschitz constant\qed

\subsection*{Proof of Lemma \ref{convex_VC_bound}}

We first show that the stage technologies $\mathcal{W}^{(s)}$ are convex under the sign restrictions for $\beta_k^{(s)}$ and $\varrho_k^{(s)}$. For the start of induction, $\mathcal{W}^{(0)}=\mathbb{R}^K$ which is convex. For the inductive step, suppose that $\mathcal{W}_k^{(s-1)}$ is convex and consider two distinct points $\mathbf{w}_1,\mathbf{w}_2\in\mathcal{W}^{(s)}$.

By construction only factors $w_{1}^{(s-1)},\dots,w_{K_{s-1}}^{(s-1)}$ produced in the $(s-1)$th layer serve as inputs in the $s$th layer, so that without loss of generality we can restrict our attention to the components $(\mathbf{w}_1^{(s-1)},\mathbf{w}_1^{(s)})$ and $(\mathbf{w}_2^{(s-1)},\mathbf{w}_2^{(s)})$ corresponding to intermediate goods produced in the $(s-1)$th and $s$th layer. Without loss of generality we assume that both points are on the boundary of $\mathcal{W}^{(s)}$, so that neither $\mathbf{w}_1\geq\mathbf{w}_2$ nor $\mathbf{w}_1\leq\mathbf{w}_2$.

By the inductive hypothesis, the components of $\lambda\mathbf{w}_1 + (1-\lambda)\mathbf{w}_2$ are in $\mathcal{W}^{(s-1)}$ for an arbitrary value of $\lambda\in[0,1]$. We can therefore verify whether $\lambda\mathbf{w}_1 + (1-\lambda)\mathbf{w}_2\in\mathcal{W}^{(s)}$ by checking whether starting at $\mathbf{w}_1$ there exists a feasible production plan to transform the vector $\lambda(\mathbf{w}_2^{(s-1)}-\mathbf{w}_1^{(s-1)})$ into the vector of quantities $\lambda(\mathbf{w}_2^{(s)}-\mathbf{w}_1^{(s)})$ for the intermediate goods at stage $s$. For $\lambda\in\{0,1\}$ this is obviously true since by assumption $\mathbf{w}_1\in\mathcal{W}^{(s)}$ and $\mathbf{w}_2\in\mathcal{W}^{(s)}$.

By inspection, the production function for each intermediate good $k$ in layer $s$ is concave in its input, where only factors $w_{1}^{(s-1)},\dots,w_{K_{s-1}}^{(s-1)}$ produced in the $(s-1)$th layer serve as inputs, so that the convex combination of the input combinations to achieve $\mathbf{w}_1^{(s)}$ and $\mathbf{w}_2^{(s)}$ achieves a vector of output quantities at stage $s$ greater than $\lambda\mathbf{w}_1 + (1-\lambda)\mathbf{w}_2$ so that indeed $\lambda\mathbf{w}_1 + (1-\lambda)\mathbf{w}_2\in\mathcal{W}^{(s)}$ for any $\lambda\in[0,1]$. Hence, $\mathcal{W}^{(s)}$ is convex for each $s$, so that the intersection $\mathcal{W}_0$ of $\mathcal{W}^{(S)}$ with any linear subspace is also convex.

We next establish the bound on the VC dimension of $\mathcal{H}$ for the case in which $\varrho_k^{(s)}>-\infty$ for all neurons. In that case, notice that the boundary of $\mathcal{W}_0$ is infinitely differentiable for each $s=0,\dots,S$, so that $\tau_{\mathbf{K},L}^S(\tilde{w};\theta)$ is also infinitely differentiable by the chain rule. By convexity of $\mathcal{W}_{0,\mathbf{K}}$ we have that all lower contour sets of $\tau_{\mathbf{K}}^S(\tilde{w};\theta)$ are fully connected, so that Theorem 2 in \cite{KMa97} applies with $B=1$, establishing the conclusion for the case in which $\varrho_k^{(s)}>-\infty$ for each neuron.

It therefore remains to be shown that the same bound applies when there may be some neurons such that $\varrho_k^{(s)}=-\infty$. To this end, consider any finite set of $Q$ points $a_1,\dots,a_Q$ that is shattered by the set of MLPs $\mathcal{H}$ allowing for unbounded values of $\varrho_k^{(s)}$. Since $Q$ is finite, we can separate the points by a distance $\delta>0$, i.e. $d(a_p,a_q)\geq\delta$ for all $p\neq q$. By continuity of neurons in each layer there must exist configurations of the MLPs with $\varrho_k^{(s)}\geq \underline{\varrho}>-\infty$ which also shatters that set. However that restricted family of MLPs again satisfies the assumptions of Theorem 2 in \cite{KMa97}, so that $Q$ must satisfy the same finite bound given in that result. In particular, $\mathcal{H}$ cannot shatter any infinite sets. Since that upper bound VC dimension for networks does not depend on a lower bound for $\varrho_k^{(s)}$, we also obtain the conclusion of the Lemma for networks when $\varrho_k^{(s)}=-\infty$ for some neurons\qed

\subsection*{Proof of Theorem \ref{conv_rate_thm}} Reviewing the proof of Theorem 1 in \cite{FLM19}, we first notice that their argument does not depend on the network being a feedforward MLP with ReLU activation functions except for the use of their Lemmas 6 and 7 which give a bound on the statistical complexity and the approximation error for the ReLU implementation of the network. In order to establish our claim, it therefore suffices to replace their Lemma 6 with Lemma \ref{convex_VC_bound} in our paper, and Propositions \ref{CES_prod_universal_prp} and \ref{CNL_nested_universal_prp}, respectively, take the place of Lemma 7 in \cite{FLM19} (which in turn corresponds to Theorem 1 in \cite{Yar17}). When applying either result, note that under Assumption \ref{data_ass}, the support of $\mathbf{z}$ is contained in a compact subset $\mathcal{C}$ of the interior of the positive orthant. The convergence rate is then obtained by substituting in the rates from these results into (A.17) in \cite{FLM19}, establishing the conclusion of the theorem\qed


\end{document}